\documentclass{article}

\usepackage[utf8]{inputenc}
\usepackage[T1]{fontenc}
\usepackage{xcolor}
\usepackage{amsmath,amssymb, amsthm}
\usepackage{algorithm}
\usepackage{algpseudocode}
\usepackage{booktabs}    
\usepackage[hidelinks]{hyperref}
\usepackage[margin=2cm]{geometry}

\usepackage{lmodern}
\usepackage{enumitem}
\usepackage{graphicx}
\usepackage[authoryear]{natbib}
\usepackage{mathtools}
\usepackage{colortbl}
\usepackage{bm}
\usepackage{accents} 
\usepackage{authblk}
\usepackage{footmisc}

\theoremstyle{plain}                      
\newtheorem{proposition}{Proposition}
\makeatletter
\@ifundefined{lemma}{\newtheorem{lemma}{Lemma}}{}
\@ifundefined{corollary}{}{}
\makeatother

\newcommand{\ABM}{{\sf ABM}}
\newcommand{\ABA}{{\sf ABA}}
\newcommand{\ceil}[1]{\left\lceil #1 \right\rceil}
\newcommand{\up}{\rule{0pt}{3ex}}
\newcommand{\NA}{---}
\newcommand{\Cpp}{C\nolinebreak\hspace{-.05em}%
  \raisebox{.4ex}{\tiny\bfseries +}\nolinebreak\hspace{-.10em}%
  \raisebox{.4ex}{\tiny\bfseries +}}

\title{Solving large-scale maximum weight non-bipartite matching in Euclidean space}

\date{}

\author[1]{P.~Baumann\thanks{Corresponding author: philipp.baumann@unibe.ch}}
\author[2]{O.~Goldschmidt}
\author[3]{D.~S.~Hochbaum}

\affil[1]{Department of Business Administration, University of Bern, Engehaldenstr.\ 4, 3012 Bern, Switzerland, philipp.baumann@unibe.ch}
\affil[2]{Riverside County Office of Education, Riverside, CA 92501, USA, goldoliv@gmail.com}
\affil[3]{Industrial Engineering and Operations Research Department, University of California, Berkeley, CA 94720, USA, dhochbaum@berkeley.edu}

\begin{document}

\maketitle
\begin{abstract}
The maximum weight non-bipartite matching problem consists of pairing the nodes of a weighted graph so that the total weight of the edges connecting paired nodes is maximized. Although the problem can be solved in polynomial time, no existing exact algorithm scales well to large graphs. Such large graphs are common in applications where the nodes are $d$-dimensional feature vectors in Euclidean space and the edge weights are pairwise distances, so that every pair of nodes is connected. We introduce \emph{Assignment-Based Matching} (\ABM), a new algorithm for non-bipartite matching in Euclidean space that builds a solution by solving one or more assignment problems. For small-sized datasets, \ABM\ delivers solutions that are only a fraction of a percent off the optimum, at much faster runtimes compared to exact algorithms. For medium- and large-sized datasets with up to six million objects, where exact algorithms fail to deliver solutions, \ABM\ produces the best known solutions compared to state-of-the-art heuristics within seconds to a few minutes.
Overall, as shown here, \ABM\ proves to be the most effective Euclidean maximum weight matching algorithm in terms of both runtime and quality of solutions.
\end{abstract}

\section{Introduction}\label{sec:intro}

Weighted matching in non-bipartite graphs is a cornerstone of combinatorial optimization. Applications of these problems occur in diverse fields including network design (see \citealt{fingerhut1997designing}), statistics (see \citealt{greevy2004optimal}), sparse linear algebra (see \citealt{schenk2006fast}), semiconductor manufacturing (see \citealt{xu2010matching}), and quantum computing (see \citealt{wu2023fusionblossom}). \citet{lu2011optimal} argue that non-bipartite matching plays an essential role in proper statistical observational studies in health and social sciences. In many applications, the nodes in the graph originate from a tabular dataset with $d$ features, leading to a graph that is embedded in the Euclidean space $\mathbb{R}^d$ (see \citealt{marcotte1991fast, greevy2004optimal}). 

We consider here the Euclidean maximum weight non-bipartite matching problem (Euclidean MWM), which is defined on graphs where the nodes are vectors in a $d$-dimensional space and the edge weights are Euclidean distances or distances induced by any other metric norm. Since these graphs are complete and all edge weights are non-negative, a maximum weight matching will be a perfect matching unless the number of nodes is odd. Here we only consider graphs with an even number of nodes, hence the Euclidean MWM is equivalent to the Euclidean maximum weight non-bipartite perfect matching problem. Note that the extension of our approaches to Euclidean MWM with an odd number of nodes is trivial. 

While the bipartite version of the maximum matching problem is efficiently solved, in theory and practice, solving the non-bipartite matching remains challenging. The maximum non-bipartite matching was shown for the first time to be solvable in polynomial time by \citet{edmonds1965paths} using the blossom algorithm. This algorithm and its extension to weighted maximum matching run in polynomial time, but are still computationally expensive. This has motivated the study of approximation algorithms and heuristics (see e.g.\ \citealt{preis1999linear, drake2003simple, pettie2004simpler, vinkemeier2005linear, papenberg2021using, wu2022solving}). The most scalable approaches for Euclidean MWM exploit the geometric structure of the problem but are restricted to specific distance metrics and/or datasets with two features (see \citealt{avis1983survey, weber1986euclidean, stanvek2026new}), whereas many applications involve considerably more features.

We propose here a new algorithm for Euclidean MWM, called \emph{Assignment-Based Matching} (\ABM), which is applicable to datasets with any number of features and to edge weights induced by any metric norm. \ABM\ constructs a matching by solving a single assignment problem, i.e., a weighted bipartite matching problem, which is much easier to solve in practice than its non-bipartite counterpart. In a variant of \ABM\ this matching is improved by solving additional assignment problems. For large-scale instances, with sizes varying from $100{,}000$ to several million objects, \ABM\ is combined with a decomposition strategy that delivers the best known solutions in seconds to minutes. 

\ABM\ is related to the recently introduced anticlustering algorithm of \citet{baumann2026fast}, called  {\em Assignment-Based Anticlustering} (\ABA). Anticlustering is an NP-hard problem that consists of partitioning a dataset of $n$ objects into $k$ equal-sized groups, anticlusters, such that the sum across all groups, of distances between objects within each group, is as large as possible (see \citealt{spath1986anticlustering}). The special case where $k=\ceil{n/2}$ is equivalent to maximum weight matching. 

We present here an extensive computational study, on nearly 100 datasets, comparing the performance of \ABM\ with that of five exact and five heuristic approaches. The results demonstrate that \ABM\ outperforms the state-of-the-art heuristics in both runtime and solution quality, and requires only a small fraction of the runtime of the exact approaches. Moreover, \ABM\ scales to sizes of datasets for which, to the best of our knowledge, no high-quality solutions have previously been reported.

The remainder of this paper is organized as follows. Section~\ref{sec:problem} formally defines the Euclidean maximum weight non-bipartite matching problem. Section~\ref{sec:literature} reviews the related literature and identifies the state-of-the-art exact and heuristic approaches. Section~\ref{sec:ABM} presents the Assignment-Based Matching (\ABM) algorithm. Section~\ref{sec:ABM-decomposition} describes a decomposition strategy that enables \ABM\ to handle large-scale instances within short runtimes, and Section~\ref{sec:ABM-improvement} proposes a procedure that iteratively improves the quality of a given solution. Section~\ref{sec:numerical_study} reports on the computational study comparing the proposed heuristics with state-of-the-art approaches. Finally, Section~\ref{sec:conclusions} concludes the paper and outlines directions for future research.

\section{Problem description}\label{sec:problem}

The Euclidean maximum weight non-bipartite matching problem is defined on a tabular dataset containing $n$ feature vectors, each $d$-dimensional. This is equivalent to a matching problem defined on a complete graph $G=(V, E)$ with $n$ nodes and edge weights $w_{ij}$ given by a distance measure between the corresponding feature vectors. The goal is to find a perfect matching of nodes which is of maximum weight (here $n$ is even). Below is a formulation of the problem as a binary program (MWM). For each edge $[i, j] \in E$, we introduce a binary decision variable $x_{ij}$, which equals $1$ if nodes $i$ and $j$ are matched, and $0$ otherwise. The objective function maximizes the total weight of the selected edges. The constraint ensures that each node $i \in V$ is matched to at most one other node.

{\renewcommand{\arraystretch}{1.5}
	\setlength{\arraycolsep}{5pt}
	\[\textup{(MWM)}\left\{\begin{array}{lll}
		\max & \displaystyle \sum_{[i, j]\in E} w_{ij}\, x_{ij} \\
		\textup{s.t.} & \displaystyle 
        \sum_{j:[i, j]\in E,\, j \neq i} x_{ij} \leq 1 & \forall i\in V  \\        
		& x_{ij} \in \{0,1\} & \forall [i,j]\in E 
	\end{array}\right.\]}

\section{Literature review}\label{sec:literature}

In this section, we review several of the existing algorithms for the maximum weight matching (MWM) problem. The exact approaches are presented in Section~\ref{sec:litexact} and the heuristic approaches in Section~\ref{sec:litheur}. 

\subsection{Exact approaches}\label{sec:litexact}

\citet{edmonds1965paths} devised the first known polynomial time algorithm for the maximum cardinality matching in general graphs, of complexity $\mathcal{O}(n^4)$, where $n$ is the number of nodes in the graph. \citet{lawler1976combinatorial} described an $\mathcal{O}(n^3)$ implementation of Edmonds' algorithm and \citet{gabow2017matching} later gave an algorithm with runtime $\mathcal{O}(m\sqrt{n})$, where $m$ is the number of edges in the graph.
Edmonds' algorithm was extended to apply to the maximum weight matching (e.g., \citealt{lawler1976combinatorial}, \citealt{edmonds2003matching}) with the complexity of $\mathcal{O}(n^3)$. For sparse graphs, $m=o(n^2)$,  \citet{galil1982priority} improved the runtime of Edmonds' algorithm to $\mathcal{O}(nm \log n)$ and \citet{gabow1990data} proposed further improvements, with a runtime $\mathcal{O}(nm + n^2 \log n)$. 

For integer edge weights, scaling algorithms have been proposed. \citet{gabow1991faster} presented a scaling algorithm that runs a modified version of Edmonds' search procedure to compute in each scaling round a near-optimal perfect matching, i.e., a perfect matching whose weight is within an additive $\mathcal{O}(n)$ of the optimum with respect to the current scaled weights. This algorithm runs in $\mathcal{O}(m\sqrt{n\,\alpha(m,n)\log n}\,\log(nN))$ time, where $N$ is the largest absolute edge weight and $\alpha$ is the inverse Ackermann function.

\citet{duan2018scaling} presented a faster scaling algorithm for maximum (or minimum) weight perfect matching that runs in $\mathcal{O}(m\sqrt{n}\log(nN))$ time. At each scaling round this algorithm computes a matching that is near-optimal and near-perfect, in that it leaves at most $\mathcal{O}(\sqrt{n})$ vertices unmatched. 

Apart from these theoretical improvements, efficient implementations of Edmonds' algorithm have been devised, most notably Blossom V \citep{kolmogorov2009blossom}, which addresses the closely related minimum cost perfect matching problem. In practice, MWM instances are 
often tackled with general-purpose solvers, either exact mixed-integer programming solvers such as Gurobi \citep{gurobi} or hybrid solvers combining local search with exact techniques such as Hexaly \citep{hexaly}. 

\subsection{Heuristic approaches}\label{sec:litheur}

Because the exact algorithms for MWM are computationally prohibitive for many practical applications, numerous heuristic approaches have been devised.

A naive greedy method, which sorts the edges by weight and repeatedly adds the heaviest remaining edge that shares no endpoint with an already selected edge, is a well-known $\frac{1}{2}$-approximation for MWM. The complexity of this greedy algorithm is dominated by the complexity of sorting the edges, $O(m\log n)$. 

There have been several improvements on the complexity of greedy: \citet{preis1999linear} presented a $\frac{1}{2}$-approximation algorithm for MWM in general graphs that runs in $\mathcal{O}(m)$ time. The idea of Preis' algorithm is to replace the heaviest edge that is used in the greedy algorithm by a \emph{locally heaviest} edge, i.e., an edge at least as heavy as all remaining edges sharing an endpoint with it. Such an edge can be found without sorting, leading to improved runtime.

Another variant of greedy that runs in linear time and delivers a $\frac{1}{2}$-approximation algorithm for MWM is by \citet{drake2003simple}. That algorithm grows node-disjoint paths: Starting from an arbitrary node, it repeatedly follows the heaviest edge incident to the current node and removes the node it leaves. The edges of each path are assigned alternately to two matchings $M_1$ and $M_2$, and the heavier of the two is returned.

A heuristic improving on the approximation factor of $\frac{1}{2}$ was described by \citet{vinkemeier2005linear}. Their algorithm delivers, for every $\epsilon > 0$, an $\frac{2}{3}- \epsilon$ approximation solution, and runs in $\mathcal{O}(m/\epsilon)$ time.

\citet{pettie2004simpler} improved the complexity dependence on $\epsilon$, from $\mathcal{O}(m/\epsilon)$ to $\mathcal{O}(m\log(1/\epsilon))$, giving both a simple randomized and a deterministic $(\frac{2}{3}-\epsilon)$-approximation algorithm.

\citet{wu2022solving} proposed a heuristic for MWM based on {\em Deep Reinforcement Learning}. A graph neural network builds the matching step by step, adding several edges at each step and discarding the edges that conflict with them. 
In their experiments this heuristic produced solutions within about 3\% of the optimum on graphs with millions of edges, while running about as fast as the greedy algorithm.

A different approach for streaming data was investigated by \citet{kale2017maximum}. The authors study maximum cardinality matching in very large unweighted graphs whose edges do not fit in memory, using the semi-streaming model of \citet{feigenbaum2005graph}. In this model, the edges arrive as a stream in an arbitrary order; the algorithm can only read the stream sequentially, possibly several times (each reading is called a \emph{pass}), and may use only $\mathcal{O}(n\,\mathrm{polylog}\,n)$ memory, where $n$ is the number of nodes. While a maximal matching, computable in one pass, is a $\frac{1}{2}$-approximation, the authors give for general graphs a two-pass $(\frac{1}{2}+\frac{1}{32})$-approximation and $(\frac{2}{3}-\epsilon)$-approximation that uses $\frac{4}{3\epsilon}$ passes and $\mathcal{O}(n\log n)$ space. 

The special case of the Euclidean MWM problem in the plane, i.e., with $d = 2$, has received considerable attention in the literature, where it is also referred to as the \emph{maximum-length matching} problem. In a 1983 survey, \citet{avis1983survey} reviews various heuristics for Euclidean MWM in the plane. 

\citet{weber1986euclidean} applied simulated annealing, based on the Metropolis algorithm, to Euclidean minimum- and maximum weight perfect matching in the plane. Starting from an initial matching, each move replaces two edges of the current matching by two other edges on the same four endpoints. Improving moves are always accepted, while worsening moves are accepted with a probability that decreases as the temperature is gradually lowered. For the maximum weight version, moves are restricted to points in roughly opposite regions of the plane, since optimal matchings of centrally symmetric point sets pass through the center. 

\citet{biniaz2025euclidean} also studied MWM in the plane. The authors make the distinction between a {\em globally maximum matching} (MWM in our parlance) and a {\em $k$-local maximum matching}. A matching is $k$-local maximum if for every subset of $k$ edges in the matching, these $k$ edges form a maximum matching for the $2k$ endpoints of those $k$ edges. They show that local maximum matchings form good approximations for global MWM. 

A new heuristic for the Euclidean MWM problem in the plane has been recently proposed, \citet{stanvek2026new}.  

We refer to this heuristic here as SA2D. SA2D builds on the maximum angular-metric traveling salesperson problem (MaxAngleTSP), which consists of finding a tour that minimizes the sum of the inner angles at its nodes rather than its length. \citet{aichholzer2017minimization} showed that for an odd number of points, an optimal MaxAngleTSP tour can be constructed in $\mathcal{O}(n \log n)$ time. For an even number of points, SA2D adds an auxiliary point, which is ignored afterwards. SA2D computes an optimal MaxAngleTSP tour and returns the heaviest of the matchings obtained by taking every second edge along this tour. In experiments on uniformly distributed random instances and TSPLIB instances with up to about 1{,}000 points, the authors observed optimality gaps of at most 1.38\%, and considerably smaller gaps on most instances. Because of this near-optimal solution quality and its $\mathcal{O}(n \log n)$ runtime, we include SA2D as a benchmark for instances in the plane. 

\citet{baumann2026fastalgorithm} have compared the performance of two anticlustering methods for MWM in Euclidean spaces of more than two dimensions. Anticlustering methods partition a set of $n$ objects into $k$ groups of equal size, called anticlusters, such that the sum of pairwise distances between objects in the same group, taken over all groups, is maximized. With $k = \ceil{n/2}$ anticlustering is equivalent to maximum weight matching. The anticlustering methods tested were the exchange-based heuristic of \citet{papenberg2021using} and an adjusted variant of the anticlustering method of \citet{baumann2026fast}. The latter algorithm forms the basis of the method developed in this paper.

\section{The Assignment-Based Matching Algorithm (\ABM)}\label{sec:ABM}

The key idea behind \ABM\ is to reduce the non-bipartite Euclidean MWM problem to a bipartite matching problem (the assignment problem), for which efficient implementations are available. The quality of the resulting matching for the original non-bipartite problem depends on how the bipartition sets ($A$ and $B$) are constructed. 
To create the bipartition, we sort the objects by their distance to the global centroid $\bm{\mu}$. The distance to the global centroid serves as a proxy for the centrality of the corresponding object. We then assign the objects to the two sets in the bipartition by alternating positions in this sorted order. Objects at adjacent positions have similar distance to the global centroid and are placed in opposite sets, so the objects in $A$ and $B$ have similar distributions of centrality. Note that centrality is a one-dimensional criterion and does not guarantee that the two sets cover the feature space equally. However, as demonstrated in our computational analysis (see Section~\ref{sec:numerical_study}), we find it to work well in practice.

For squared Euclidean distances, the centrality sorting admits an additional motivation. Let $r_i = \|\bm{x}_i - \bm{\mu}\|$ denote the distance of object $i$ to the global centroid. By the triangle inequality, $\|\bm{x}_i - \bm{x}_j\| \le r_i + r_j$, so the objective value of any perfect matching $M$ satisfies

\begin{equation}\label{eq:centroid_bound}
  \sum_{\{i,j\} \in M} \|\bm{x}_i - \bm{x}_j\|^2
  \;\le\; \sum_{\{i,j\} \in M} (r_i + r_j)^2
  \;=\; \sum_{i=1}^{n} r_i^2 \;+\; 2 \sum_{\{i,j\} \in M} r_i r_j .
\end{equation}

The first term on the right-hand side does not depend on $M$. The second term is largest when objects with similar distances to the centroid are paired: for $r_1 \ge r_2 \ge r_3 \ge r_4$, the pairing $\{1,2\},\{3,4\}$ yields $r_1 r_2 + r_3 r_4$, which is at least as large as $r_1 r_3 + r_2 r_4$ and $r_1 r_4 + r_2 r_3$. Moreover, the triangle inequality holds with equality if and only if $\bm{\mu}$ lies on the line segment between $\bm{x}_i$ and $\bm{x}_j$. Bound~\eqref{eq:centroid_bound} thus describes an ideal pair: two objects with
similar distance to the global centroid that lie on opposite sides of it. The alternating selection strategy places objects at adjacent positions in the sorted order, which have similar distances to the centroid, in opposite sets, and leaves it to the assignment problem to select partners on opposite sides of the centroid.
Objects two positions apart, however, also have similar distances to the centroid but are placed in the same set, so an ideal pair may be excluded.

\begin{figure}
    \centering
    \includegraphics[width=\textwidth]{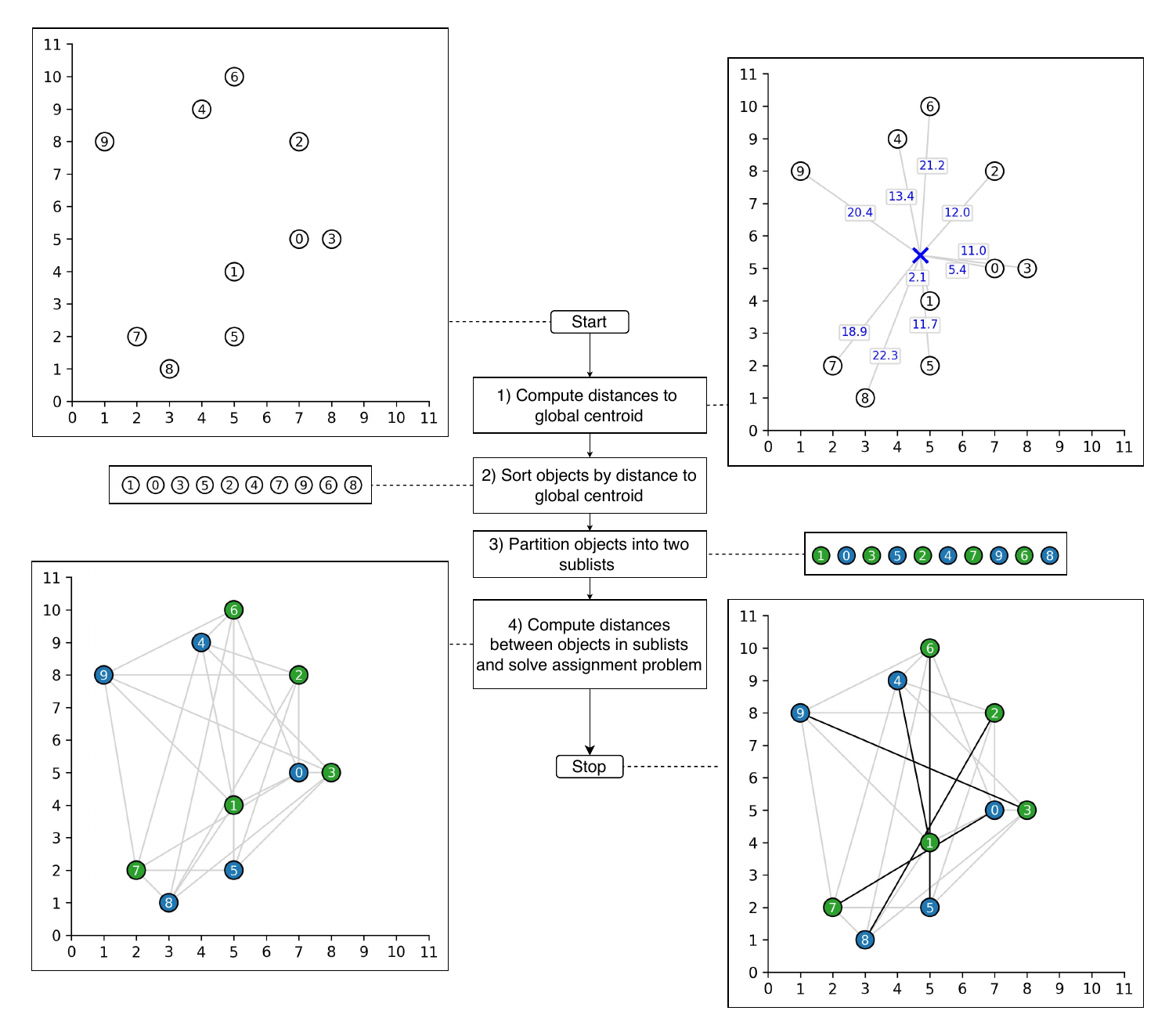}
    \caption{Application of  \ABM\ to the illustrative example.}
    \label{fig:flowchart}
\end{figure}

Algorithm~\ref{alg:abm} gives the pseudocode. The algorithm proceeds in four steps. First, it computes the distance of each object to the global centroid $\bm{\mu} = \frac{1}{n}\sum_{i=1}^n\bm{x}_i$ of the dataset. Second, it sorts the objects by increasing distance and stores their indices in a list $N$. Third, it partitions $N$ into two sets: set $A$ holds the indices at odd positions of $N$, and set $B$ those at even positions. Fourth, it computes the distances between the objects in $A$ and those in $B$ and solves the maximum weight assignment problem matching each object of $A$ to one of $B$. The resulting pairing is the output of the algorithm. Figure~\ref{fig:flowchart} illustrates the procedure for an example with $n=10$ objects. 

\begin{algorithm}[ht]
\caption{The Assignment-Based Matching Algorithm (\ABM)}
\begin{algorithmic}[1]
\Require Set of objects: ${\mathcal{X}}=\{\bm{x}_1, \ldots, \bm{x}_n\}$, distance function: DIST()
\Ensure Set of pairs: $\mathcal{P}=\{(p_1,q_1), \ldots, (p_{n/2},q_{n/2})\}$
\State $\text{distances} \gets \text{empty vector of size } n$
\State $\bm{\mu} \gets \frac{1}{n}\sum_{i=1}^n\bm{x}_i$ \Comment{Compute global centroid}
\For{$i=1$ \textbf{to} $n$}
    \State $\text{distances}[i] \gets \text{DIST}(\bm{x}_i, \bm{\mu})$ \Comment{Compute distance to global centroid}
\EndFor
\State $N \gets \text{argsort}(\text{distances})$ \Comment{Sort objects and store their indices in $N$}
\State $A \gets \text{indices at odd positions of $N$}$
\State $B \gets \text{indices at even positions of $N$}$
\State $\text{weightmatrix} \gets \text{empty matrix of size } |A| \times |B|$
\For{$i=1$ \textbf{to} $|A|$} 
    \For{$j=1$ \textbf{to} $|B|$}
        \State $\text{weightmatrix}[i][j] \gets \text{DIST}(\bm{x}_{A[i]}, \bm{x}_{B[j]})$ \Comment{Compute distance}
    \EndFor
\EndFor
\State $Q \gets \text{SOLVE\_MAX\_WEIGHT\_ASSIGNMENT}(\text{weightmatrix})$ 
\State $\mathcal{P} \gets \emptyset$ \Comment{Initialize set of pairs}
\For{$(u, v) \in Q$}
    \State $\mathcal{P} \gets \mathcal{P} \cup \{(A[u], B[v])\}$ \Comment{Map back to object indices}
\EndFor
\end{algorithmic}
\label{alg:abm}
\end{algorithm}

As mentioned in the introduction, the \ABM~algorithm for Euclidean MWM can use any norm, $L_q$ for $q \geq 1$. To solve the assignment problem, we use the \texttt{linear\_sum\_assignment} function provided by the Python SciPy library, which implements the modified Jonker--Volgenant algorithm described in \citet{crouse2016implementing}.

The complexity of the \ABM~algorithm for Euclidean MWM is dominated by solving the maximum weight assignment problem. Computing the global centroid and the distance of each object to it takes $O(nd)$ operations, and sorting the objects by centrality takes $O(n \log n)$. Constructing the weight matrix requires all pairwise distances between the two partite sets, at a cost of $O(n^2 d)$. Solving the maximum weight assignment problem on this matrix with the Jonker--Volgenant algorithm takes $O(n^3)$ operations, since each partite set contains $n/2$ objects. The overall complexity is therefore $O(n^3 + n^2 d)$, which reduces to $O(n^3)$ whenever $d = O(n)$, as is typically the case.

\begin{proposition}\label{prop:half}
For squared Euclidean distances, the matching returned by \ABM\ has an objective value of at least half the optimal objective value. This holds for any partition of the objects into two sets of equal size.
\end{proposition}

The proof, given in the Appendix, compares both solutions with the total scatter $\mathrm{SSE} = \sum_{i=1}^{n} \|\bm{x}_i - \bm{\mu}\|^2$. Every perfect matching has an objective value of at most $2\,\mathrm{SSE}$, whereas the
optimal assignment between $A$ and $B$ is at least as good as a uniformly random assignment, whose expected objective value is at least $\mathrm{SSE}$. The bound is tight as shown in the Appendix. In practice, however, the gaps are far smaller (see Section~\ref{sec:numerical_study}).

\section{The Assignment-Based Matching Algorithm with Decomposition}\label{sec:ABM-decomposition}

Applying \ABM\ to $n$ objects requires solving an assignment problem of size $n/2$, which costs $O(n^3)$ and becomes prohibitive as $n$ grows. We therefore propose a scalable variant of \ABM\ based on a decomposition strategy. It takes as input the dataset $\mathcal{X}$, an assignment problem size $s$, and a distance function DIST(). The idea is to partition the objects into $m = \lceil n/(2s) \rceil$ batches of at most $2s$ objects each. The objects of each batch are then matched separately using \ABM. Finally, the resulting pairs are combined to give the final solution. This decomposition strategy reduces the cost to $O(m s^3) = O(n s^2)$, which is linear in $n$ for a fixed assignment problem size $s$. The subproblems are independent and can be solved in parallel. 

Algorithm~\ref{alg:decomposition} gives the pseudocode. The algorithm proceeds in three steps. First, it randomly shuffles the object indices and stores them in a list $N$. Second, it splits $N$ into $m$ batches of size at most $2s$. Third, it applies \ABM\ to the objects in each batch, yielding $s$ pairs per batch. The pairs from all batches form the matching $\mathcal{P}$, which is the output of the algorithm.

\begin{algorithm}[ht]
\caption{Assignment-Based Matching Algorithm with Decomposition (\ABM-d)}
\label{alg:decomposition}
\begin{algorithmic}[1]
\Require Set of objects: ${\mathcal{X}}=\{\bm{x}_1, \ldots, \bm{x}_n\}$, assignment problem size: $s$, distance function: DIST()
\Ensure Set of pairs: $\mathcal{P}=\{(p_1,q_1), \ldots, (p_{n/2},q_{n/2})\}$
\State $N \gets$ random permutation of $(1, \ldots, n)$ \Comment{Randomly shuffle the objects}
\State $m \gets \lceil n/(2s) \rceil$ \Comment{Number of subproblems}
\State Split $N$ into batches $N_1, \ldots, N_m$ with only the last one of size possibly less than $2s$
\State $\mathcal{P} \gets \emptyset$ \Comment{Initialize set of pairs}
\For{$b=1$ \textbf{to} $m$}
     \State Call \ABM\ with $\{\bm{x}_i : i \in N_b\}$ and DIST, and store the resulting pairs in~$Q$
    \For{$(u,v) \in Q$}
        \State $\mathcal{P} \gets \mathcal{P} \cup \{(N_b[u], N_b[v])\}$ \Comment{Map back to object indices}
    \EndFor
\EndFor
\end{algorithmic}
\end{algorithm}

\section{Assignment-based improvement procedure}\label{sec:ABM-improvement}

The core idea of the assignment-based matching algorithm (\ABM) can also be used to improve an existing solution. In this section, we propose an assignment-based improvement procedure that iteratively refines a given matching. 

The procedure takes as input the dataset $\mathcal{X}$, a set of pairs $\mathcal{P}$ representing the matching to be improved, the number of improvement iterations $T$, the size $s$ of each assignment problem, and a distance function DIST(). In each iteration, the procedure randomly assigns one member of each pair to set $A$ and the other to set $B$, and then solves one or multiple assignment problems to rematch the objects in $A$ with those in $B$. Since the current matching is always one of the possible solutions, the objective function value can only stay the same or improve.

A single assignment problem has a time complexity of $O(n^3)$, which is prohibitive for large $n$. We therefore propose to split $A$ and $B$ into $m = \lceil n/(2s) \rceil$ batches and solve one assignment problem per batch, reducing the cost to $O(m s^3)$. Objects can then only be rematched within their batch. To let objects meet new partners from other batches, the order of the pairs is randomly permuted at the start of every iteration, so that the batches are composed differently each time. Because the pairs are permuted as units, both members of every current pair remain in the same batch. The current matching restricted to each batch is therefore a feasible solution of that batch's assignment problem, and the objective function value cannot decrease. The batches are independent and can be solved in parallel. Setting $s = n/2$ recovers the unbatched procedure, in which case the permutation has no effect.

Algorithm~\ref{alg:abi} gives the pseudocode. Each iteration proceeds in three steps. First, each pair $(p_c, q_c) \in \mathcal{P}$ is split at random, so that $A$ and $B$ contain one member of every pair. Second, both sets are divided at the same positions into $m = \lceil n/(2s) \rceil$ batches with only the last ones of size possibly less than $s$. Third, for each batch $b=1,\ldots,m$ the distances between the objects in $A_b$ and those in $B_b$ are computed and the maximum weight assignment problem is solved. The pairs from all batches form the new matching $\mathcal{P}$, which is the input for the next iteration.

\begin{algorithm}
\caption{Assignment-based improvement procedure}
\label{alg:abi}
\begin{algorithmic}[1]
\Require Set of objects: ${\mathcal{X}}=\{\bm{x}_1, \ldots, \bm{x}_n\}$, set of pairs: $\mathcal{P}=\{(p_1,q_1), \ldots, (p_{n/2},q_{n/2})\}$, number of iterations: $T$, assignment problem size: $s$, distance function: DIST()
\Ensure Set of pairs: $\mathcal{P}=\{(p_1,q_1), \ldots, (p_{n/2},q_{n/2})\}$
\For{$t=1$ \textbf{to} $T$} \Comment{Perform $T$ improvement iterations}
    \State Randomly permute the order of the pairs in $\mathcal{P}$ 
    \State $A \gets \text{empty vector of size } n/2$, \quad $B \gets \text{empty vector of size } n/2$ 
    \For{$c=1$ \textbf{to} $n/2$}
        \State draw $r$ uniformly at random from $\{0,1\}$ 
        \If{$r = 0$} \Comment{Assign one member of each pair to A and the other to B}
            \State $A[c] \gets p_c$, \quad $B[c] \gets q_c$
        \Else
            \State $A[c] \gets q_c$, \quad $B[c] \gets p_c$
        \EndIf
    \EndFor
    \State $m \gets \lceil n/(2s) \rceil$ \Comment{Number of batches, given $s$}
    \State Split $A, B$ into batches $A_1, \ldots, A_m$ and $B_1, \ldots, B_m$ at the same positions
    \State $\mathcal{P} \gets \emptyset$ \Comment{Initialize set of pairs}
    \For{$b=1$ \textbf{to} $m$}
        \State $\text{weightmatrix} \gets \text{empty matrix of size } |A_b| \times |B_b|$
        \For{$i=1$ \textbf{to} $|A_b|$}
            \For{$j=1$ \textbf{to} $|B_b|$}
                \State $\text{weightmatrix}[i][j] \gets \text{DIST}(\bm{x}_{A_b[i]}, \bm{x}_{B_b[j]})$ \Comment{Compute distance}
            \EndFor
        \EndFor

        \State $Q \gets \text{SOLVE\_MAX\_WEIGHT\_ASSIGNMENT}(\text{weightmatrix})$
        \For{$(u, v) \in Q$}
            \State $\mathcal{P} \gets \mathcal{P} \cup \{(A_b[u], B_b[v])\}$ \Comment{Append new pair}
        \EndFor
    \EndFor
\EndFor
\end{algorithmic}
\end{algorithm}

Since the improvement procedure starts from the solution of \ABM\ and never
decreases the objective value, the guarantee of Proposition~\ref{prop:half}
carries over.

\section{Computational experiments and analysis}\label{sec:numerical_study}

We compare the performance of \ABM\ with that of state-of-the-art algorithms in five computational experiments. Section~\ref{sec:results-algorithms} describes the exact and heuristic algorithms included in the comparison. For instances not solved to optimality, we assess solution quality using upper bounds on the optimal objective value; Section~\ref{sec:results-bounds} explains how we compute these bounds. Sections~\ref{sec:results-small} to~\ref{sec:results-large} present the first three experiments, in which we analyze the performance of the algorithms on small-, medium-, and large-sized problem instances, respectively. Section~\ref{sec:results-ablation} presents an ablation study, in which we analyze the impact of different components of \ABM\ on the solution quality. Finally, in Section~\ref{sec:results-tsp}, we compare \ABM\ to the algorithm of \citet{stanvek2026new} which is only applicable to datasets where the objects are points in the Euclidean plane and the edge weights correspond to Euclidean distances. \ABM\ is implemented in Python and our code is publicly available on GitHub (see \url{https://github.com/phil85/abm}). We ran all experiments on a Mac Studio with an M4 Max chip (16 CPU cores) and 128 GB of RAM. All reported runtimes are wall-clock times.

\subsection{Tested algorithms} \label{sec:results-algorithms}

We compare the performance of \ABM\ to five exact and five heuristic approaches. Table~\ref{tbl:tested-algorithms} provides an overview of the tested algorithms, along with the abbreviations used to refer to them throughout the remainder of this section. We next describe the five exact and the five heuristic benchmark approaches.

\begin{table}
\setlength{\tabcolsep}{15pt}
\caption{Overview of tested algorithms}\label{tbl:tested-algorithms}
\footnotesize
\begin{tabular}{llll}
\toprule
Abbr.\ & Approach & Code & Source \\
\midrule
\multicolumn{4}{l}{\textit{Exact approaches}} \\
\addlinespace[2pt]
GRB & Gurobi 13.0 & C & \href{www.gurobi.com}{\color{blue}gurobi.com} \\
HEX-B & Hexaly 15.0 with binary variables & \Cpp{} & \href{www.hexaly.com}{\color{blue}hexaly.com} \\
HEX-L & Hexaly 15.0 with list variable & \Cpp{} & \href{www.hexaly.com}{\color{blue}hexaly.com} \\
NWX & \texttt{max\_weight\_matching()} of NetworkX & Python & \href{https://networkx.org}{\color{blue}networkx.org} \\
BLV & Blossom V version 2.05 of \citet{kolmogorov2009blossom} & \Cpp{} & \href{https://pub.ista.ac.at/~vnk/software.html#BLOSSOM5}{\color{blue}Source code} \\ \midrule
\multicolumn{4}{l}{\textit{Heuristic approaches}} \\
\addlinespace[2pt]
AC & \texttt{anticlustering()} of anticlust & C & \href{https://github.com/m-Py/anticlust/tree/main}{\color{blue}GitHub}  \\
FAC-X & \texttt{fast\_anticlustering()} of anticlust with X random exchange partners & C & \href{https://github.com/m-Py/anticlust/tree/main}{\color{blue}GitHub}  \\
GRD & The greedy heuristic as described in \citet{avis1983survey} & Python & \href{https://github.com/phil85/abm}{\color{blue}GitHub}  \\
RND & Random matching & Python & \href{https://github.com/phil85/abm}{\color{blue}GitHub}  \\
SA2D & Algorithm for Euclidean MWM in the plane & Python & On request$^{1)}$ \\
\midrule
\multicolumn{4}{l}{\textit{Algorithms introduced in this paper}} \\
\addlinespace[2pt]
\ABM & The Assignment-Based Matching Algorithm (\ABM) & Python & \href{https://github.com/phil85/abm}{\color{blue}GitHub}  \\
\ABM-iX & \ABM\ with X improvement iterations & Python & \href{https://github.com/phil85/abm}{\color{blue}GitHub}  \\
\ABM-dX & \ABM\ with decomposition using assignment problem size $X$ & Python & \href{https://github.com/phil85/abm}{\color{blue}GitHub}  \\
{\sf rABM} & \ABM\ with randomly generated sets $A$ and $B$ & Python & \href{https://github.com/phil85/abm}{\color{blue}GitHub}  \\

\bottomrule
\end{tabular}
\up $^{1)}$Code not publicly available; obtained from the authors.
\end{table}

The first two exact approaches solve the binary programming formulation MWM (see Section~\ref{sec:problem}): GRB uses the Gurobi solver (see \citealt{gurobi}) and HEX-B uses the Hexaly solver (see \citealt{hexaly}). Both require all pairwise distances to be computed in advance, and the number of decision variables grows quadratically with the number of objects. The third exact approach (HEX-L) avoids this quadratic growth by solving an alternative formulation with Hexaly. Besides binary, integer, and continuous variables, Hexaly offers \emph{list} decision variables, which represent an ordered sequence of distinct elements. HEX-L uses a single list variable that must contain all $n$ objects, so that every value of the variable is a permutation of the objects. The objects at positions $2i-1$ and $2i$, for $i=1,\dots,n/2$, form the $i$-th pair; for example, the list $(4,1,6,2,3,5)$ yields the pairs $\{4,1\}$, $\{6,2\}$, and $\{3,5\}$. The objective maximizes the sum of the within-pair distances, where each distance is computed directly from the feature values of the two objects rather than taken from a precomputed $n \times n$ distance matrix. Consequently, the memory requirement of HEX-L grows only linearly with the size of the data matrix. Hexaly explores the space of permutations mainly by local search, i.e., by repeatedly modifying the current list (e.g., by moving or swapping elements) and keeping improving changes, while separately computing bounds to certify optimality. Since every permutation corresponds to a feasible perfect matching, HEX-L quickly finds feasible solutions even for very large instances. However, because the objective is a nonlinear function of the list positions, the bounds that Hexaly derives for this formulation are weak; in our experiments, HEX-L did not prove optimality for any instance. The fourth exact approach is the routine \texttt{max\_weight\_matching()} from the Python library NetworkX (see \citealt{hagberg2008networkx}), which implements the blossom algorithm of \citet{edmonds1965paths}. The fifth exact approach (BLV) is Blossom V of \citet{kolmogorov2009blossom}, an efficient \Cpp{} implementation of Edmonds' algorithm for the minimum weight perfect matching problem. To find a maximum weight perfect matching with Blossom V, we set the weight of edge $[i,j]$ to $d_{\max} - d_{ij}$, where $d_{\max}$ is the largest pairwise distance. As Blossom V requires integer edge weights, we scale these weights linearly such that the largest weight equals $10^8$, which keeps them well within the range of 32-bit integers, and round them to the nearest integer. The source code of Blossom V is publicly available at \url{https://pub.ista.ac.at/~vnk/software.html#BLOSSOM5}. We additionally tested the Fusion Blossom algorithm of \citet{wu2023fusionblossom}, which is another implementation of the algorithm by \citet{edmonds1965paths} originally developed for quantum error correction. Since it requires edge weights to be even integers, we transformed our data accordingly before applying it. However, the Blossom V and the NetworkX routines were consistently faster, so we do not present the results of Fusion Blossom here.

The first heuristic approach is the routine \texttt{anticlustering()} from the R package anticlust of \citet{papenberg2021using}. We use it with \texttt{objective = "variance"} and set the number of anticlusters to the number of pairs required. We chose the variance objective because it is equivalent to the objective of the MWM (see Section~\ref{sec:problem}) formulation when squared Euclidean distances are used as edge weights. The other control parameters were left at their default values. The second heuristic is the routine \texttt{fast\_anticlustering()} from the R package anticlust of \citet{papenberg2021using}. We applied it with different numbers of randomly generated exchange partners. The source code of the anticlust package is publicly available at \url{https://github.com/m-Py/anticlust/tree/main}. The third heuristic approach is the greedy heuristic as described in \citet{avis1983survey}. It sorts all edges by decreasing weight and iteratively selects the heaviest available edge that does not conflict with an already matched node until a perfect matching is formed. We implemented the greedy heuristic in Python using the NetworkX package (see \citealt{hagberg2008networkx}). The code is available at \url{https://github.com/phil85/abm}. The fourth heuristic is random matching. This heuristic first randomly permutes the objects. Then, the object at position $i=1,\ldots,n$ of the permuted order is assigned to pair $c=(i \bmod (n/2)) + 1$. This approach eliminates the need to explicitly enumerate the edges. We implemented random matching in Python. The code is available at \url{https://github.com/phil85/abm}. The fifth heuristic is the recently published algorithm of~\citet{stanvek2026new} for maximum weight matching in the Euclidean plane. Since it is only applicable to datasets with exactly two features, we refer to it as SA2D. We are grateful to the authors for kindly sharing their Python implementation of the algorithm with us.

Three heuristics are deterministic (\ABM, GRD, and SA2D), while the others are randomized and require a random seed (\ABM-iX, \ABM-dX, AC, FAC-X, RND). We applied each randomized heuristic three times per instance, each time with a different random seed, and report the average objective function value and runtime in the results tables. Across all experiments, we imposed a time limit of two hours per run. Approaches that did not return a solution within this limit were terminated. In the results tables, a dash (\NA) indicates that no solution was returned within the time limit. 

GRB, HEX-B and HEX-L were run with the default thread settings of Gurobi and Hexaly and therefore used all 16 cores. All other approaches run on a single thread; in particular, the \texttt{linear\_sum\_assignment} routine of SciPy used by \ABM\ is single-threaded. The only exception is \ABM-d, whose subproblems are independent: in Sections~\ref{sec:results-medium}, \ref{sec:results-large}, and \ref{sec:results-tsp} we solve them in parallel on 16 cores. 

We also considered including the reinforcement learning heuristic of \citet{wu2022solving} as a benchmark heuristic. In preliminary experiments with the pre-trained models available on  \url{https://github.com/HowePa/L2M} it turned out to be substantially less effective on complete Euclidean graphs than on the sparse graphs it was originally trained and evaluated on. Retraining it on our instances was beyond the computational resources available to us, as the required GPU infrastructure was not at our disposal. We therefore excluded it from our analysis.

\subsection{Upper bounds}\label{sec:results-bounds}

For some medium- and all large-sized datasets, the optimal objective function values are unknown. To assess the quality of the solutions on these datasets, we compute upper bounds on the optimal objective function value. All bounds rely on the following observation. Let $y_1,\ldots,y_n$ be real numbers satisfying
\begin{equation}\label{eq:dual}
y_i + y_j \geq w_{ij} \qquad \forall\, [i,j] \in E.
\end{equation}
Because an optimal perfect matching (with objective function value $z_{\text{opt}}$)  covers every node exactly once, inequality~\eqref{eq:dual} implies $z_{\text{opt}} \leq \sum_{i=1}^n y_i$. We construct $y$ in two ways, which yield the \emph{star bound} and the \emph{farthest-neighbor bound}.

The \emph{star bound} sets $y_i = 2\,\|\bm{x}_i - \bm{\mu}\|^2$ for squared Euclidean distances. This choice satisfies \eqref{eq:dual} because, by the parallelogram law, $\|\bm{u} - \bm{v}\|^2 + \|\bm{u} + \bm{v}\|^2 = 2\|\bm{u}\|^2 + 2\|\bm{v}\|^2$ for all vectors $\bm{u}$ and $\bm{v}$, and hence $\|\bm{u} - \bm{v}\|^2 \leq 2\|\bm{u}\|^2 + 2\|\bm{v}\|^2$. Setting $\bm{u} = \bm{x}_i - \bm{\mu}$ and $\bm{v} = \bm{x}_j - \bm{\mu}$ gives $w_{ij} \leq y_i + y_j$. For the Euclidean and the Manhattan distance, the triangle inequality shows that $y_i = \text{DIST}(\bm{x}_i, \bm{c})$ satisfies \eqref{eq:dual} for any center $\bm{c}$. We use the geometric median and the coordinate-wise median, respectively, which minimize the resulting bound. The star bound requires $O(nd)$ operations for squared Euclidean and Manhattan distances. For Euclidean distances, we approximate the geometric median with Weiszfeld's algorithm \citep{weiszfeld2009point}, which requires $O(nd)$ operations per iteration; since any center yields a valid bound, an approximate geometric median suffices.

The \emph{farthest-neighbor bound} sets $y_i = \frac{1}{2}\max_{j \neq i} w_{ij}$, which satisfies \eqref{eq:dual} for arbitrary edge weights and requires $O(n^2 d)$ operations. The two bounds complement each other: the star bound is typically tighter for low-dimensional data, and the farthest-neighbor bound for high-dimensional data.

Finally, we tighten the better of the two bounds by repeatedly updating all $y_i$ simultaneously according to
\[
y_i \leftarrow \tfrac{1}{2}\Bigl(y_i + \max_{j \neq i}\,(w_{ij} - y_j)\Bigr).
\]
If $y$ satisfies \eqref{eq:dual}, then so does the updated vector, and its sum does not increase. Each iteration requires $O(n^2 d)$ operations, which we implement with blockwise matrix products. We stop when an iteration reduces the bound by less than 0.001\%. Let $\mathrm{UB}$ denote the resulting bound and $z_{\text{best}}$ the best known objective value. We report the upper bound gap (in \%) as  $\frac{z_{\text{best}} - \mathrm{UB}}{z_{\text{best}}} \times 100.$ In our experiments, the computation of the upper bounds for the large-sized datasets is extremely time-consuming. For example, it takes multiple days (about 89.7 h) to compute the upper bound of the largest dataset with over 6 million objects.

\subsection{Results on small datasets}\label{sec:results-small}

In our first experiment, we apply the algorithms to ten small datasets for which most exact approaches can deliver provably optimal solutions within the two-hour time limit. This allows us to assess which heuristics are able to find optimal or near-optimal solutions. Table~\ref{tbl:datasets_small} lists, for each of the ten datasets, its name, the number of objects ($n$), the number of features ($d$), the data type, and the source. Each dataset is a random subset of $2{,}000$ objects drawn from the corresponding original dataset available on the \href{https://archive.ics.uci.edu}{UCI ML Repository \color{blue}(https://archive.ics.uci.edu)}.  Before sampling, we followed the steps described in \citet{baumann2026fast} to preprocess the original datasets.

\begin{table}
    \centering
	\footnotesize
	\caption{Small datasets\label{tbl:datasets_small}}    
	\setlength{\tabcolsep}{6pt}
	\begin{tabular}{lrrll}
\toprule 
 Dataset & $n$ & $d$ & Type & Source \\ \midrule 
Pulsar-2k & 2,000 & 8 & Tabular & \href{https://archive.ics.uci.edu/dataset/372/htru2}{\color{blue}UCI ML Repository} \\
Abalone-2k & 2,000 & 10 & Tabular & \href{https://archive.ics.uci.edu/dataset/1/abalone}{\color{blue}UCI ML Repository} \\
Electric-2k & 2,000 & 12 & Tabular & \href{https://archive.ics.uci.edu/dataset/471/electrical+grid+stability+simulated+data}{\color{blue}UCI ML Repository} \\
Facebook-2k & 2,000 & 13 & Tabular & \href{https://archive.ics.uci.edu/dataset/488/facebook+live+sellers+in+thailand}{\color{blue}UCI ML Repository} \\
Frogs-2k & 2,000 & 22 & Tabular & \href{https://archive.ics.uci.edu/dataset/406/anuran+calls+mfccs}{\color{blue}UCI ML Repository} \\
Creditcard-2k & 2,000 & 24 & Tabular & \href{https://archive.ics.uci.edu/dataset/350/default+of+credit+card+clients}{\color{blue}UCI ML Repository} \\
Travel-2k & 2,000 & 24 & Tabular & \href{https://archive.ics.uci.edu/dataset/485/tarvel+review+ratings}{\color{blue}UCI ML Repository} \\
Bank-2k & 2,000 & 53 & Tabular & \href{https://archive.ics.uci.edu/dataset/222/bank+marketing}{\color{blue}UCI ML Repository} \\
Plants-2k & 2,000 & 70 & Tabular & \href{https://archive.ics.uci.edu/dataset/180/plants}{\color{blue}UCI ML Repository} \\
Adult-2k & 2,000 & 110 & Tabular & \href{https://archive.ics.uci.edu/dataset/2/adult}{\color{blue}UCI ML Repository} \\
\bottomrule
\end{tabular}

\end{table}

We applied our assignment-based matching algorithm with two settings: the standard algorithm (\ABM), and with 5 improvement iterations (\ABM-i5). The five exact approaches (GRB, HEX-B, HEX-L, NWX, and BLV) and three of the heuristics (AC, GRD, and RND) serve as benchmark methods. For such small instances, it is not necessary to use \ABM\ with decomposition (\ABM-dX) or to use \texttt{fast\_anticlustering()} (FAC-X). The performance metrics considered are solution quality (objective function value) and runtime.

Table~\ref{tbl:ofv_small} reports the solution quality achieved by the different approaches. The first three columns give, for each dataset, its name, the number of objects ($n$), and the number of features ($d$). The fourth column reports the optimal objective function value for each instance. The fifth column reports the relative gap between the upper bound (UB) and the optimal objective function value $z_{\text{opt}}$, computed as $\frac{z_{\text{opt}} - \textup{UB}}{z_{\text{opt}}} \times 100$. The remaining columns report, for each approach, the percentage deviation from the optimal objective function value, computed as $\frac{z_{\text{opt}} - z_{\text{approach}}}{z_{\text{opt}}} \times 100$. The last row states the average deviation for each approach. 

\begin{table}
\scriptsize
\caption{Objective function values for small datasets\label{tbl:ofv_small}}
\setlength{\tabcolsep}{5pt}
\begin{tabular}{lrrrrrrrrrrrrrr}
\toprule
 & & & & & \multicolumn{10}{c}{Deviation from optimal objective (\%)} \\
\cmidrule(lr){6-15}
 & & & \multicolumn{1}{l}{Optimal} & \multicolumn{1}{l}{Upper} & & & & & & & & & & \\
 & & & \multicolumn{1}{l}{objective} & \multicolumn{1}{l}{bound} & & & & & & & & & & \\
Dataset & $n$ & $d$ & \multicolumn{1}{l}{($z_{\text{opt}}$)} & \multicolumn{1}{l}{gap (\%)} & GRB & HEX-B & HEX-L & NWX & BLV & \textbf{ABM} & \textbf{\textbf{ABM}-i5} & AC & GRD & RND \\
\midrule
Pulsar-2k & 2,000 & 8 & 20,119.86 & -4.24 & 0.00 & 0.00 & 0.06 & 0.00 & 0.00 & 0.24 & 0.03 & 0.36 & 3.89 & 36.82 \\
Abalone-2k & 2,000 & 10 & 36,229.74 & -0.82 & 0.00 & 0.00 & 0.02 & 0.00 & 0.00 & 0.07 & 0.01 & 0.20 & 0.73 & 43.70 \\
Electric-2k & 2,000 & 12 & 43,820.25 & -0.35 & 0.00 & 0.00 & 1.17 & 0.00 & 0.00 & 1.70 & 0.21 & 3.12 & 2.28 & 45.06 \\
Facebook-2k & 2,000 & 13 & 28,936.39 & -7.40 & 0.00 & 0.00 & 0.00 & 0.00 & 0.00 & 0.00 & 0.00 & 0.01 & 5.81 & 19.52 \\
Frogs-2k & 2,000 & 22 & 81,833.59 & -4.89 & 0.00 & 0.00 & 0.13 & 0.00 & 0.00 & 0.30 & 0.03 & 0.71 & 3.84 & 37.26 \\
Creditcard-2k & 2,000 & 24 & 69,188.42 & -1.92 & 0.00 & 0.00 & 0.22 & 0.00 & 0.00 & 0.35 & 0.04 & 0.78 & 6.97 & 32.15 \\
Travel-2k & 2,000 & 24 & 75,166.04 & -1.50 & 0.00 & 0.00 & 0.34 & 0.00 & 0.00 & 0.39 & 0.04 & 1.20 & 4.64 & 32.26 \\
Bank-2k & 2,000 & 53 & 61,541.42 & -3.60 & 0.00 & 0.00 & 0.02 & 0.00 & 0.00 & 0.07 & 0.01 & 0.11 & 2.49 & 18.83 \\
Plants-2k & 2,000 & 70 & 19,284.00 & -0.73 & 0.00 & 0.00 & 0.00 & 0.00 & 0.00 & 0.00 & 0.00 & 0.00 & 4.82 & 16.13 \\
Adult-2k & 2,000 & 110 & 235,968.78 & -0.30 & 0.00 & 0.00 & 0.05 & 0.00 & 0.00 & 0.05 & 0.01 & 0.12 & 1.85 & 9.84 \\
\midrule Average &  &  &  & &  0.00 & 0.00 & 0.20 & 0.00 & 0.00 & 0.32 & 0.04 & 0.66 & 3.73 & 29.16 \\
\bottomrule
\end{tabular}

\up \ABM\ and \ABM-i5 are introduced in this paper. The abbreviations are explained in Table~\ref{tbl:tested-algorithms}. The upper bound (UB) gap is $\frac{z_{\text{opt}} - \textup{UB}}{z_{\text{opt}}} \times 100$.
\end{table}

The exact approaches GRB, HEX-B, NWX, and BLV find certified optimal solutions within the time limit for all instances. The exact approach HEX-L fails to prove optimality within the time limit for all instances and returns a suboptimal solution for all except two instances. Our heuristic \ABM\ matches or outperforms every benchmark heuristic (AC, GRD, and RND) on each instance, achieving an average deviation of just 0.32\% compared to 0.66\%, 3.73\%, and 29.16\% for AC, GRD, and RND, respectively. \ABM-i5, which is \ABM\ with 5 improvement iterations, further enhances solution quality. With an average deviation of 0.04\% and a maximum deviation of 0.21\%, it finds near-optimal solutions on every instance. On the small datasets, where the optimal objective values are known, the upper bounds exceed the optimal objective values by only 0.30\% to 7.40\% (2.58\% on average), which shows that the bound is reasonably tight and serves as a meaningful reference for the medium and large datasets.

Table~\ref{tbl:cpu_small} reports the runtime (in seconds) required by each approach. Among the exact approaches, GRB is the fastest, taking on average 5.80 seconds to prove optimality, compared to 33.92 seconds for BLV, 429.19 seconds for HEX-B, and 1,742.60 seconds for NWX. As mentioned above, HEX-L cannot prove optimality within the two-hour time limit and therefore runs for the full 7,200 seconds on every instance. Among the heuristics, only RND is faster than \ABM\ and \ABM-i5, but at the cost of vastly inferior solution quality (see Table~\ref{tbl:ofv_small}). AC and GRD are both slower than \ABM\ and \ABM-i5 despite offering worse solutions. \ABM\ takes just 0.16 seconds on average, roughly 35 times faster than GRB, the fastest exact approach, while still achieving near-optimal solution quality (see Table~\ref{tbl:ofv_small}). Even \ABM-i5, which performs additional improvement iterations to further refine solution quality, requires only 0.93 seconds on average, still about 6 times faster than GRB and orders of magnitude faster than HEX-B and NWX. Notably, the runtime of AC increases with the number of features~($d$).

\begin{table}
\scriptsize
\caption{Runtimes for small datasets in seconds\label{tbl:cpu_small}}
\setlength{\tabcolsep}{7.3pt}
\begin{tabular}{lrrrrrrrrrrrr}
\toprule
 & & & \multicolumn{10}{c}{Runtime (s)} \\
\cmidrule(lr){4-13}
Dataset & $n$ & $d$ & GRB & HEX-B & HEX-L & NWX & BLV & \textbf{ABM} & \textbf{\textbf{ABM}-i5} & AC & GRD & RND \\
\midrule
Pulsar-2k & 2,000 & 8 & 5.96 & 450.17 & 7,200.23 & 1,729.14 & 26.17 & \bfseries 0.16 & 0.90 & 6.77 & 1.77 &$<$0.01\\
Abalone-2k & 2,000 & 10 & 6.23 & 452.84 & 7,199.89 & 1,727.65 & 69.87 & \bfseries 0.16 & 0.93 & 7.60 & 1.76 &$<$0.01\\
Electric-2k & 2,000 & 12 & 4.88 & 423.64 & 7,200.13 & 1,631.54 & 5.41 & \bfseries 0.07 & 0.38 & 9.20 & 1.77 &$<$0.01\\
Facebook-2k & 2,000 & 13 & 6.55 & 454.14 & 7,200.02 & 2,260.27 & 97.77 & \bfseries 0.22 & 1.51 & 10.38 & 1.67 &$<$0.01\\
Frogs-2k & 2,000 & 22 & 5.86 & 415.50 & 7,199.88 & 1,867.74 & 50.30 & \bfseries 0.16 & 0.92 & 16.04 & 1.65 &$<$0.01\\
Creditcard-2k & 2,000 & 24 & 6.20 & 437.55 & 7,200.00 & 1,793.01 & 13.86 & \bfseries 0.16 & 0.95 & 16.65 & 1.77 &$<$0.01\\
Travel-2k & 2,000 & 24 & 5.70 & 439.71 & 7,199.81 & 1,700.48 & 20.87 & \bfseries 0.14 & 0.77 & 16.33 & 1.70 &$<$0.01\\
Bank-2k & 2,000 & 53 & 6.62 & 440.74 & 7,200.10 & 2,062.01 & 34.46 & \bfseries 0.20 & 1.30 & 34.33 & 1.78 &$<$0.01\\
Plants-2k & 2,000 & 70 & 3.96 & 336.08 & 7,200.92 & 529.61 & 2.69 & \bfseries 0.04 & 0.20 & 42.30 & 1.42 &$<$0.01\\
Adult-2k & 2,000 & 110 & 6.01 & 441.54 & 7,201.75 & 2,124.53 & 17.77 & \bfseries 0.28 & 1.56 & 67.15 & 1.82 &$<$0.01\\
\midrule Average &  &  & 5.80 & 429.19 & 7,200.27 & 1,742.60 & 33.92 & \bfseries 0.16 & 0.94 & 22.67 & 1.71 &$<$0.01\\
\bottomrule
\end{tabular}

\up \ABM\ and \ABM-i5 are introduced in this paper. The lowest runtime per dataset is shown in bold (excluding RND).
\end{table}

To assess whether our findings depend on the choice of distance measure, we repeated this experiment using the Manhattan ($L_1$) and the Euclidean ($L_2$) distance instead of the squared Euclidean distance. We excluded HEX-L from these experiments because it performed worse than HEX-B on the small instances. Overall, the results were qualitatively the same, and all conclusions drawn in this section remain valid. For brevity, we do not report these results here, but they are available in our GitHub repository (see \url{https://github.com/phil85/abm}). 

\subsection{Results on medium-sized datasets}\label{sec:results-medium}

In our second experiment, we apply the algorithms to thirteen medium-sized datasets with up to 60,000 objects and 3,072 features. Our goal here is to assess the scalability of the heuristic approaches. Table~\ref{tbl:datasets_medium} lists, for each of the thirteen datasets, its name, the number of objects ($n$), the number of features ($d$), the data type, and the source. Unlike in the first experiment, we use the full datasets rather than a sampled subset. We followed the steps described in \citet{baumann2026fast} to preprocess the datasets. In addition, for datasets containing an odd number of objects (rows), we removed the last object to ensure an even count.

\begin{table}
	\footnotesize
    \centering
	\caption{Medium-sized datasets\label{tbl:datasets_medium}}
	\setlength{\tabcolsep}{6pt}
	\begin{tabular}{lrrll}
\toprule 
 Dataset & $n$ & $d$ & Type & Source \\ \midrule 
Abalone & 4,176 & 10 & Tabular & \href{https://archive.ics.uci.edu/dataset/1/abalone}{\color{blue}UCI ML Repository} \\
Travel & 5,454 & 24 & Tabular & \href{https://archive.ics.uci.edu/dataset/485/tarvel+review+ratings}{\color{blue}UCI ML Repository} \\
Facebook & 7,050 & 13 & Tabular & \href{https://archive.ics.uci.edu/dataset/488/facebook+live+sellers+in+thailand}{\color{blue}UCI ML Repository} \\
Frogs & 7,194 & 22 & Tabular & \href{https://archive.ics.uci.edu/dataset/406/anuran+calls+mfccs}{\color{blue}UCI ML Repository} \\
Electric & 10,000 & 12 & Tabular & \href{https://archive.ics.uci.edu/dataset/471/electrical+grid+stability+simulated+data}{\color{blue}UCI ML Repository} \\
Npi & 10,440 & 40 & Tabular & \href{https://openpsychometrics.org/_rawdata/}{\color{blue}Online personality tests} \\
Pulsar & 17,898 & 8 & Tabular & \href{https://archive.ics.uci.edu/dataset/372/htru2}{\color{blue}UCI ML Repository} \\
Creditcard & 30,000 & 24 & Tabular & \href{https://archive.ics.uci.edu/dataset/350/default+of+credit+card+clients}{\color{blue}UCI ML Repository} \\
Adult & 32,560 & 110 & Tabular & \href{https://archive.ics.uci.edu/dataset/2/adult}{\color{blue}UCI ML Repository} \\
Plants & 34,780 & 70 & Tabular & \href{https://archive.ics.uci.edu/dataset/180/plants}{\color{blue}UCI ML Repository} \\
Bank & 45,210 & 53 & Tabular & \href{https://archive.ics.uci.edu/dataset/222/bank+marketing}{\color{blue}UCI ML Repository} \\
Cifar10 & 50,000 & 3,072 & Image & \href{https://www.cs.toronto.edu/~kriz/cifar.html}{\color{blue}UCI ML Repository} \\
Mnist & 60,000 & 784 & Image & \href{https://www.kaggle.com/datasets/hojjatk/mnist-dataset?select=t10k-labels.idx1-ubyte}{\color{blue}Kaggle} \\
\bottomrule
\end{tabular}

\end{table}

We applied our assignment-based matching algorithm with four settings: the standard algorithm (\ABM), the standard algorithm with one and two improvement iterations (\ABM-i1, \ABM-i2), and the standard algorithm with decomposition using an assignment problem size of $s=5{,}000$ (\ABM-d5000). For the two variants with improvement iterations (\ABM-i1 and \ABM-i2), we used an assignment problem size of $s=n/2$. The exact approaches GRB, HEX-B, HEX-L, BLV, and the four heuristics AC, FAC-500, GRD, and RND serve as benchmark methods. NWX was not able to return a solution within the time limit for any of the medium-sized instances. For this reason, we do not include it in the results tables here. As before, the performance metrics considered are solution quality (objective function value) and runtime.

Table~\ref{tbl:ofv_medium} reports the solution quality achieved by the different approaches. The table is structured analogously to the corresponding one for the small datasets. Since the exact approaches did not provide a certified optimal solution for each dataset, the fourth column of the table reports the best objective function value found across all approaches for each instance, rather than the proven optimal value. The fifth column reports the relative gap between the upper bound (UB) and the best objective function value $z_{\text{best}}$, computed as $\frac{z_{\text{best}} - \textup{UB}}{z_{\text{best}}} \times 100$. The remaining columns report, for each approach, the percentage deviation from the best objective function value, computed as $\%DEV = \frac{z_{\text{best}} - z_{\text{approach}}}{z_{\text{best}}} \times 100$. The last row states the average deviation for each approach. We report this average only for approaches that produced a solution for every instance.

\begin{table}
\tiny
\caption{Objective values for medium-sized datasets\label{tbl:ofv_medium}}
\setlength{\tabcolsep}{3.6pt}
\begin{tabular}{lrrrrrrrrrrrrrrrr}
\toprule
 & & & & & \multicolumn{12}{c}{Deviation from best objective (\%)} \\
\cmidrule(lr){6-17}
 & & & \multicolumn{1}{l}{Best} & \multicolumn{1}{l}{Upper} & & & & & & & & & & & & \\
 & & & \multicolumn{1}{l}{objective} & \multicolumn{1}{l}{bound} & & & & & & & & & & & & \\
Dataset & $n$ & $d$ & \multicolumn{1}{l}{($z_{\text{best}}$)} & \multicolumn{1}{l}{gap (\%)} & GRB & HEX-B & HEX-L & BLV & \textbf{ABM} & \textbf{\textbf{ABM}-i1} & \textbf{\textbf{ABM}-i2} & \textbf{\textbf{ABM}-d5000} & AC & FAC-500 & GRD & RND \\
\midrule
Abalone & 4,176 & 10 & 74,817.50 & -0.87 & 0.00 & 0.00 & 0.07 & 0.00 & 0.06 & 0.03 & 0.02 & 0.06 & 0.19 & 0.48 & 0.77 & 44.20 \\
Travel & 5,454 & 24 & 193,609.69 & -2.08 & 0.00 & 0.44 & 0.79 & 0.00 & 0.27 & 0.11 & 0.06 & 0.27 & 0.93 & 2.52 & 4.37 & 32.06 \\
Facebook & 7,050 & 13 & 105,777.75 & -5.47 & 0.00 & 0.00 & 0.00 & 0.00 & 0.00 & 0.00 & 0.00 & 0.00 & 0.00 & 0.00 & 8.75 & 12.91 \\
Frogs & 7,194 & 22 & 255,910.88 & -2.62 & 0.00 & 0.84 & 1.11 & 0.00 & 0.32 & 0.15 & 0.09 & 0.32 & 0.95 & 2.39 & 6.23 & 38.33 \\
Electric & 10,000 & 12 & 225,348.23 & -0.22 & 0.00 & \NA & 5.01 & 0.00 & 1.06 & 0.52 & 0.33 & 1.06 & 2.72 & 8.25 & 2.01 & 46.86 \\
Npi & 10,440 & 40 & 134,708.00 & -21.24 & 0.00 & \NA & 5.09 & 0.00 & 0.00 & 0.00 & 0.00 & 0.26 & 2.04 & 5.61 & 7.18 & 35.23 \\
Pulsar & 17,898 & 8 & 218,298.67 & -5.34 & 0.00 & \NA & 1.05 & \NA & 0.07 & 0.03 & 0.02 & 0.17 & 0.20 & 1.01 & 4.42 & 34.16 \\
Creditcard & 30,000 & 24 & 1,057,515.25 & -2.98 & \NA & \NA & 5.27 & \NA & 0.09 & 0.02 & 0.00 & 0.38 & \NA & 2.36 & 6.31 & 31.97 \\
Adult & 32,560 & 110 & 3,967,987.47 & -0.30 & \NA & \NA & 1.89 & \NA & 0.01 & 0.00 & 0.00 & 0.05 & \NA & \NA & 1.41 & 9.73 \\
Plants & 34,780 & 70 & 302,103.00 & -1.07 & \NA & \NA & 0.07 & \NA & 0.00 & 0.00 & 0.00 & 0.00 & \NA & \NA & 3.64 & 14.51 \\
Bank & 45,210 & 53 & 2,942,646.16 & -2.92 & \NA & \NA & 3.92 & \NA & 0.02 & 0.00 & 0.00 & 0.13 & \NA & \NA & \NA & 18.53 \\
Cifar10 & 50,000 & 3,072 & 15,766,598.45 & -0.80 & \NA & \NA & 35.79 & \NA & 0.43 & 0.12 & 0.00 & 2.14 & \NA & \NA & \NA & 39.52 \\
Mnist & 60,000 & 784 & 4,383,206.85 & -1.78 & \NA & \NA & 21.36 & \NA & 0.24 & 0.06 & 0.00 & 1.38 & \NA & \NA & \NA & 27.87 \\
\midrule Average &  &  &  & &  \NA & \NA & 6.26 & \NA & 0.20 & 0.08 & 0.04 & 0.48 & \NA & \NA & \NA & 29.68 \\
\bottomrule
\end{tabular}

\scriptsize
\up \ABM, \ABM-i1, \ABM-i2 and \ABM-d5000 are introduced in this paper. The abbreviations are explained in Table~\ref{tbl:tested-algorithms}. The upper bound (UB) gap is $\frac{z_{\text{best}} - \textup{UB}}{z_{\text{best}}} \times 100$.
\end{table}

\ABM\ again matches or outperforms the benchmark heuristics (AC, FAC-500, GRD, and RND) on every dataset and finds optimal or near-optimal solutions with an average deviation of 0.20\% and a maximum deviation of 1.05\%. The proposed assignment-based improvement iterations further enhance solution quality: \ABM-i1 reduces the average deviation to 0.08\% (maximum 0.52\%), and \ABM-i2 to just 0.04\% (maximum 0.32\%). \ABM-d5000 delivers the same solution quality as \ABM\ on all datasets with fewer than $10{,}000$ objects and performs slightly worse than \ABM\ for the larger ones. \ABM-d5000 still considerably outperforms the benchmark heuristics AC, FAC-500, GRD, and RND. As \ABM-i1 and \ABM-i2 demonstrate, the assignment-based improvement procedure could narrow this gap further. For larger datasets, the improvement procedure can be kept tractable by using a moderate assignment problem size, e.g., $s = 5000$. We did not apply the improvement procedure to \ABM-d5000, since that variant is intended to demonstrate scalability and already achieves high solution quality. Among the benchmark heuristics, only RND produces solutions for the three largest datasets. AC, FAC-500, and GRD all fail to do so. Interestingly, GRD finds solutions for larger instances than AC and FAC-500, even though it sorts all pairwise distances while AC and FAC-500 operate directly on the dataset rather than the distance matrix. However, when AC and FAC-500 do find solutions, they deliver the best solution quality among the four benchmark heuristics AC, FAC-500, GRD, and RND. RND delivers by far the worst solution quality, with an average deviation of 29.68\%. On the seven smallest datasets, GRB outperforms the other two solver-based exact approaches, HEX-B and HEX-L. HEX-L is the only exact approach that returns a solution for all datasets within the time limit. However, its solutions are consistently worse than those of \ABM, and on the two largest datasets, its solutions are only moderately better than those obtained with RND. For the seven datasets solved by GRB (Abalone to Pulsar), the best objective value is optimal, so the UB gap measures the tightness of the upper bound. The bound lies within 3\% of the optimum on four of these datasets but is considerably weaker on Facebook (5.47\%), Pulsar (5.34\%) and Npi (21.24\%). For the six largest datasets, which have no GRB results, the UB gap bounds the optimality gap of the best solutions, all of which are attained by \ABM-i2: these solutions are within 0.30\% to 2.98\% of the upper bound. Since the upper bound is not tight even where the optimum is known, the true optimality gaps are likely smaller.

\begin{table}
\tiny
\caption{Runtimes for medium-sized datasets in seconds\label{tbl:cpu_medium}}
\setlength{\tabcolsep}{3.8pt}
\begin{tabular}{lrrrrrrrrrrrrrr}
\toprule
 & & & \multicolumn{12}{c}{Runtime (s)} \\
\cmidrule(lr){4-15}
Dataset & $n$ & $d$ & GRB & HEX-B & HEX-L & BLV & \textbf{ABM} & \textbf{\textbf{ABM}-i1} & \textbf{\textbf{ABM}-i2} & \textbf{\textbf{ABM}-d5000} & AC & FAC-500 & GRD & RND \\
\midrule
Abalone & 4,176 & 10 & 34.47 & 4,619.92 & 7,200.69 & 879.83 & 1.26 & 2.43 & 3.59 & 1.30 & 69.17 & \bfseries 0.33 & 8.57 &$<$0.01\\
Travel & 5,454 & 24 & 52.47 & 7,322.19 & 7,200.82 & 603.96 & 2.50 & 4.74 & 6.90 & 2.53 & 283.23 & \bfseries 0.46 & 14.93 &$<$0.01\\
Facebook & 7,050 & 13 & 98.17 & 11,213.84 & 7,201.05 & 5,770.09 & 8.46 & 17.73 & 27.45 & 8.55 & 417.22 & \bfseries 0.48 & 25.46 &$<$0.01\\
Frogs & 7,194 & 22 & 107.53 & 7,876.87 & 7,200.91 & 3,499.80 & 5.44 & 10.35 & 15.12 & 5.49 & 621.44 & \bfseries 0.56 & 27.65 &$<$0.01\\
Electric & 10,000 & 12 & 171.24 & \NA & 7,201.06 & 525.30 & 5.68 & 10.32 & 14.48 & 5.75 & 1,064.19 & \bfseries 0.74 & 57.42 &$<$0.01\\
Npi & 10,440 & 40 & 313.31 & \NA & 7,202.94 & 1,802.36 & 6.24 & 11.22 & 15.86 & 5.67 & 3,039.43 & \bfseries 1.00 & 43.04 &$<$0.01\\
Pulsar & 17,898 & 8 & 1,094.56 & \NA & 7,201.04 & \NA & 99.47 & 189.26 & 277.99 & 19.57 & 4,446.20 & \bfseries 1.22 & 195.99 &$<$0.01\\
Creditcard & 30,000 & 24 & \NA & \NA & 7,205.98 & \NA & 445.08 & 894.32 & 1,339.09 & 20.16 & \NA & \bfseries 2.20 & 783.79 &$<$0.01\\
Adult & 32,560 & 110 & \NA & \NA & 7,224.94 & \NA & 1,121.07 & 2,221.40 & 3,321.13 & \bfseries 38.10 & \NA & \NA & 2,968.28 &$<$0.01\\
Plants & 34,780 & 70 & \NA & \NA & 7,215.93 & \NA & 90.63 & 122.71 & 154.62 & \bfseries 2.86 & \NA & \NA & 1,958.71 &$<$0.01\\
Bank & 45,210 & 53 & \NA & \NA & 7,216.05 & \NA & 2,155.42 & 3,922.25 & 5,687.82 & \bfseries 28.46 & \NA & \NA & \NA &$<$0.01\\
Cifar10 & 50,000 & 3,072 & \NA & \NA & 8,237.64 & \NA & 1,196.64 & 2,250.01 & 3,241.17 & \bfseries 30.89 & \NA & \NA & \NA &$<$0.01\\
Mnist & 60,000 & 784 & \NA & \NA & 7,497.86 & \NA & 1,327.40 & 2,598.93 & 3,815.22 & \bfseries 17.56 & \NA & \NA & \NA &$<$0.01\\
\midrule Average &  &  & \NA & \NA & 7,308.22 & \NA & 497.33 & 942.74 & 1,378.50 & \bfseries 14.38 & \NA & \NA & \NA &$<$0.01\\
\bottomrule
\end{tabular}

\scriptsize
\up \ABM, \ABM-i1, \ABM-i2 and \ABM-d5000 are introduced in this paper. The lowest runtime per dataset is shown in bold (excluding RND).
\end{table}

Table~\ref{tbl:cpu_medium} reports the runtime (in seconds) required by each approach. The runtime of the exact approach (GRB) grows quickly with $n$ and it fails to find a feasible solution within the time limit for datasets with $n\geq30{,}000$. Among the heuristics, only the assignment-based heuristics (\ABM, \ABM-i1, \ABM-i2, and \ABM-d5000) as well as random matching (RND) find solutions for all datasets within the time limit. RND remains, as before, by far the fastest approach, but at the cost of very poor solution quality (see Table~\ref{tbl:ofv_medium}). The runtime of \ABM\ increases with increasing number of objects ($n$) as the size of the assignment problems to be solved increase with $n$. \ABM-i1 and \ABM-i2 require roughly twice and three times as long as \ABM, respectively, consistent with the additional number of assignment problems solved. \ABM-d5000 offers a promising trade-off between solution quality and scalability. By decomposing large instances into subproblems of size $s=5{,}000$, runtime is low and relatively stable, averaging 14.38 seconds across all 13 datasets. This is roughly 35 times faster than \ABM\ on average and 75 times faster on the {\sf Bank} dataset specifically. The runtime of AC grows rapidly with $n$ and AC fails to return a solution within the time limit for datasets larger than {\sf Pulsar}. FAC-500 shows a similar pattern, failing to return a solution within the time limit beyond {\sf Creditcard}. GRD is slower than \ABM\ and both of its improvement variants (\ABM-i1 and \ABM-i2) on most datasets and fails to complete on the three largest instances ({\sf Bank}, {\sf Cifar10}, and {\sf Mnist}). 

As with the small datasets, we repeated this experiment with the Manhattan ($L_1$) and Euclidean ($L_2$) distances. We excluded \texttt{fast\_anticlustering} from these experiments because it optimizes the $k$-means objective 
which is specific to squared Euclidean distances. The results are qualitatively similar, and our conclusions remain unchanged (see \url{https://github.com/phil85/abm}). Remarkably, with the Manhattan distance, \ABM-i2 finds provably optimal solutions for {\sf Adult}, {\sf Plants} and {\sf Bank}, for which no GRB results are available: their objective values equal the upper bound. To the best of our knowledge, these are the largest instances of the MWM problem with Manhattan distances that have been solved to proven optimality.

\subsection{Results on large datasets}\label{sec:results-large}

In our third experiment, we apply the most scalable algorithms (\ABM-dX, FAC-5, HEX-L, and RND) to eight large datasets with up to 6 million objects and up to 3,072 features. Our goal here is to push the scalability limits of the assignment-based heuristics and to assess the impact of the assignment problem size $s$ used for the decomposition on runtime and solution quality. Table~\ref{tbl:datasets_large} lists, for each of the eight datasets, its name, the number of objects ($n$), the number of features ($d$), the data type, and the source. As in the second experiment, we use the full datasets and followed the steps described in \citet{baumann2026fast} to preprocess them. Again, for datasets containing an odd number of objects (rows), we removed the last object to ensure an even count.

\begin{table}
    \centering
	\scriptsize
	\caption{Large datasets\label{tbl:datasets_large}}
	\setlength{\tabcolsep}{6pt}
	\begin{tabular}{lrrll}
\toprule 
 Dataset & $n$ & $d$ & Type & Source \\ \midrule 
Survival & 110,204 & 4 & Tabular & \href{https://archive.ics.uci.edu/dataset/827/sepsis+survival+minimal+clinical+records}{\color{blue}UCI ML Repository} \\
Diabetes & 253,680 & 22 & Tabular & \href{https://archive.ics.uci.edu/dataset/891/cdc+diabetes+health+indicators}{\color{blue}UCI ML Repository} \\
Music & 515,344 & 91 & Tabular & \href{https://archive.ics.uci.edu/dataset/203/yearpredictionmsd}{\color{blue}Kaggle} \\
Covtype & 581,012 & 55 & Tabular & \href{https://archive.ics.uci.edu/dataset/31/covertype}{\color{blue}UCI ML Repository} \\
Imagenet8 & 1,281,166 & 192 & Image & \href{https://www.image-net.org/index.php}{\color{blue}ImageNet Image Database} \\
Imagenet32 & 1,281,166 & 3,072 & Image & \href{https://www.image-net.org/index.php}{\color{blue}ImageNet Image Database} \\
Census & 2,458,284 & 68 & Tabular & \href{https://archive.ics.uci.edu/dataset/116/us+census+data+1990}{\color{blue}UCI ML Repository} \\
Finance & 6,362,620 & 12 & Tabular & \href{https://www.kaggle.com/datasets/sriharshaeedala/financial-fraud-detection-dataset}{\color{blue}Kaggle} \\
\bottomrule
\end{tabular}

\end{table}

We applied \ABM-dX with five different assignment problem sizes: $1{,}000$, $2{,}000$, $\ldots$, $5{,}000$. The only three benchmark approaches that were able to produce at least one solution for the large datasets within the time limit are HEX-L, FAC-5 and RND. All other exact approaches (GRB, HEX-B, NWX, and BLV) and heuristic approaches (AC, GRD) failed to deliver a solution within the time limit. GRD ran out of memory because it requires the complete graph as input. As before, the performance metrics considered are solution quality (objective function value) and runtime. 

\begin{table}
\tiny
\caption{Objective values for large datasets\label{tbl:ofv_large}}
\setlength{\tabcolsep}{3.9pt}
\begin{tabular}{lrrrrrrrrrrrr}
\toprule
 & & & & & \multicolumn{8}{c}{Deviation from best objective (\%)} \\
\cmidrule(lr){6-13}
 & & & \multicolumn{1}{l}{Best} & \multicolumn{1}{l}{Upper} & & & & & & & & \\
 & & & \multicolumn{1}{l}{objective} & \multicolumn{1}{l}{bound} & & & & & & & & \\
Dataset & $n$ & $d$ & \multicolumn{1}{l}{($z_{\text{best}}$)} & \multicolumn{1}{l}{gap (\%)} & HEX-L & \textbf{\textbf{ABM}-d1000} & \textbf{\textbf{ABM}-d2000} & \textbf{\textbf{ABM}-d3000} & \textbf{\textbf{ABM}-d4000} & \textbf{\textbf{ABM}-d5000} & FAC-5 & RND \\
\midrule
Survival & 110,204 & 4 & 663,840.25 & -6.63 & 2.43 & 0.03 & 0.01 & 0.00 & 0.00 & 0.00 & 4.60 & 33.53 \\
Diabetes & 253,680 & 22 & 7,979,938.03 & -4.38 & 20.31 & 0.73 & 0.33 & 0.16 & 0.06 & 0.00 & \NA & 30.05 \\
Music & 515,344 & 91 & 61,814,684.86 & -5.16 & 23.34 & 2.02 & 1.10 & 0.59 & 0.25 & 0.00 & \NA & 24.15 \\
Covtype & 581,012 & 55 & 37,666,117.82 & -1.63 & 14.24 & 0.41 & 0.20 & 0.10 & 0.04 & 0.00 & \NA & 15.16 \\
Imagenet8 & 1,281,166 & 192 & 23,109,921.67 & -5.22 & 42.13 & 2.28 & 1.21 & 0.65 & 0.27 & 0.00 & \NA & 42.36 \\
Imagenet32 & 1,281,166 & 3,072 & 428,597,408.33 & -5.39 & \NA & 2.24 & 1.19 & 0.64 & 0.27 & 0.00 & \NA & 38.01 \\
Census & 2,458,284 & 68 & 219,878,062.39 & -2.47 & 23.92 & 0.32 & 0.14 & 0.07 & 0.03 & 0.00 & \NA & 23.98 \\
Finance & 6,362,620 & 12 & 88,425,514.07 & -2.77 & 13.59 & 0.00 & 0.00 & 0.00 & 0.00 & 0.00 & \NA & 13.65 \\
\midrule Average &  &  &  & &  \NA & 1.00 & 0.52 & 0.28 & 0.12 & 0.00 & \NA & 27.61 \\
\bottomrule
\end{tabular}

\scriptsize
\up The approaches \ABM-dX are introduced in this paper. The abbreviations are explained in Table~\ref{tbl:tested-algorithms}. The upper bound (UB) gap is $\frac{z_{\text{best}} - \textup{UB}}{z_{\text{best}}} \times 100$.
\end{table}

Table~\ref{tbl:ofv_large} reports the solution quality achieved by the different approaches. The table is structured analogously to the corresponding one for the medium-sized datasets. FAC-5 is only able to return a solution for the smallest dataset ({\sf Survival}) within the time limit. We found that FAC-5 does not scale well with the number of features $d$: {\sf Survival} has only $d=4$ features, which allowed FAC-5 to find a solution, while for some of the medium-sized datasets with more features it failed to return a solution within the time limit. All assignment-based matching heuristics (\ABM-dX) turn out to be highly scalable: they produce a solution for every dataset, including the very high-dimensional image datasets and the million-scale ones. Solution quality consistently increases with increasing size of the assignment problem $s$. The random matching heuristic RND is also highly scalable, but produces solutions of considerably worse quality. The exact approach HEX-L finds feasible solutions for all datasets except the high-dimensional {\sf Imagenet32} dataset. However, even though it uses the full two-hour time limit, its solutions are considerably worse than those found by the assignment-based heuristics (\ABM-d$X$) in much shorter runtimes. The gaps from the upper bounds to the best objectives range from 1.63\% ({\sf Covtype}) to 6.63\% ({\sf Survival}), averaging 4.21\%. The best solutions, all obtained by \ABM-d5000, are therefore provably within 6.63\% of optimality, and likely closer, since the bound need not be tight. These solutions could be improved further by applying the assignment-based improvement procedure (e.g., with an assignment problem size $s=5000$), but we omitted this step given the already small optimality gaps. Note that computing the upper bounds for such large datasets is time-consuming. For example, to compute the upper bound of {\sf Imagenet32}, 12 tightening iterations were required, taking 323,075 seconds (about 89.7 h). 

\begin{table}
\tiny
\caption{Runtimes for large datasets in seconds\label{tbl:cpu_large}}
\setlength{\tabcolsep}{7.5pt}
\begin{tabular}{lrrrrrrrrrr}
\toprule
 & & & \multicolumn{8}{c}{Runtime (s)} \\
\cmidrule(lr){4-11}
Dataset & $n$ & $d$ & HEX-L & \textbf{\textbf{ABM}-d1000} & \textbf{\textbf{ABM}-d2000} & \textbf{\textbf{ABM}-d3000} & \textbf{\textbf{ABM}-d4000} & \textbf{\textbf{ABM}-d5000} & FAC-5 & RND \\
\midrule
Survival & 110,204 & 4 & 7,204.08 & \bfseries 0.66 & 2.23 & 5.86 & 8.13 & 13.94 & 2.89 & 0.01 \\
Diabetes & 253,680 & 22 & 7,241.48 & \bfseries 1.69 & 6.24 & 14.58 & 24.85 & 41.67 & \NA & 0.02 \\
Music & 515,344 & 91 & 7,524.23 & \bfseries 3.35 & 10.92 & 21.66 & 34.61 & 53.22 & \NA & 0.04 \\
Covtype & 581,012 & 55 & 7,417.57 & \bfseries 5.49 & 21.34 & 43.84 & 81.94 & 126.70 & \NA & 0.04 \\
Imagenet8 & 1,281,166 & 192 & 8,869.02 & \bfseries 8.47 & 27.97 & 55.34 & 95.51 & 143.84 & \NA & 0.10 \\
Imagenet32 & 1,281,166 & 3,072 & \NA & \bfseries 58.02 & 127.72 & 210.64 & 292.38 & 386.26 & \NA & 0.10 \\
Census & 2,458,284 & 68 & 8,371.66 & \bfseries 26.55 & 109.67 & 241.25 & 435.05 & 702.47 & \NA & 0.19 \\
Finance & 6,362,620 & 12 & 7,983.34 & \bfseries 70.86 & 283.36 & 616.68 & 1,157.23 & 1,773.48 & \NA & 0.49 \\
\midrule Average &  &  & \NA & \bfseries 21.89 & 73.68 & 151.23 & 266.21 & 405.20 & \NA & 0.12 \\
\bottomrule
\end{tabular}

\scriptsize
\up The \ABM-dX approaches are introduced in this paper. The lowest runtime per dataset is shown in bold (excluding RND).
\end{table}

Table~\ref{tbl:cpu_large} reports the runtime (in seconds) required by each approach. Remarkably, \ABM-d1000 finds high-quality solutions for every dataset in a few seconds, or a bit over a minute at most. Runtime then increases with increasing size of the assignment problem $s$, making parameter $s$ an effective control parameter for managing the trade-off between runtime and solution quality. RND is again by far the fastest heuristic, but as in the previous experiments, it delivers poor solution quality. HEX-L, as mentioned above, always uses the full two-hour time limit.

As for the small- and medium-sized datasets, we repeated this experiment with the Manhattan ($L_1$) and Euclidean ($L_2$) distances. Again, the results were qualitatively the same, and our conclusions remain valid (see \url{https://github.com/phil85/abm}). 

\subsection{Ablation study}\label{sec:results-ablation}

In this experiment, we use twelve synthetic datasets to systematically study the impact of different components of \ABM\ on solution quality. We generated these datasets using the \texttt{make\_blobs()} routine of the scikit-learn Python package, varying the number of objects ($n$) and the number of features ($d$) across datasets. For each dataset, we set the parameter \textit{centers} to ten and the parameter \textit{cluster\_std} to 0.1 to give the data a high level of structure, with ten tight, well-separated clusters. 

We applied four variants of the assignment-based matching algorithm: the standard \ABM\ algorithm; a variant referred to as {\sf rABM} that generates the bipartite sets A and B randomly rather than from sorted distances; and two variants with 5 and 10 improvement iterations, respectively (\ABM-i5 and \ABM-i10). Both of these variants use an assignment problem size $s=n/2$. This allows us to isolate the contribution of each component to the overall improvement, namely, the sorting-based generation of A and B, the assignment-problem, and the improvement iterations. As reference points, we also applied the exact approach GRB and the heuristics AC, GRD, and RND. As before, the performance metrics considered are solution quality (objective function value) and runtime.

Tables~\ref{tbl:ofv_synthetic} and \ref{tbl:cpu_large_synthetic} report the solution quality and the runtimes, respectively. We can draw the following conclusions from the tables: 

\begin{itemize}
    \item Even when the bipartite sets A and B are generated randomly, the assignment-based matching heuristic ({\sf rABM}) outperforms the benchmark heuristics (AC, GRD, and RND) on average, with an average percentage deviation from the optimal objective function values of only 0.44\%.
    \item The sorting-based generation of the bipartite sets A and B consistently delivers higher solution quality than the random-based generation. This is evident from \ABM\ delivering noticeably better solution quality than {\sf rABM} on every dataset: {\sf rABM} fails to find any optimal solution, whereas \ABM\ finds six.
    \item The improvement iterations are effective and increase runtime in a predictable manner. For the synthetic datasets considered here, only 10 iterations were needed to find an optimal solution for every dataset.
    \item All assignment-based variants outperformed the other heuristics on average in terms of solution quality. Moreover, they scale better with the number of objects $n$ than AC and GRD; only RND, which yields much poorer solutions, scales better.    
\end{itemize}

\begin{table}
\tiny
\caption{Objective values for synthetic datasets\label{tbl:ofv_synthetic}}
\setlength{\tabcolsep}{8.3pt}
\begin{tabular}{lrrrrrrrrrrrr}
\toprule
 & & & & & \multicolumn{8}{c}{Deviation from optimal objective (\%)} \\
\cmidrule(lr){6-13}
 & & & \multicolumn{1}{l}{Optimal} & \multicolumn{1}{l}{Upper} & & & & & & & & \\
 & & & \multicolumn{1}{l}{objective} & \multicolumn{1}{l}{bound} & & & & & & & & \\
Dataset & $n$ & $d$ & \multicolumn{1}{l}{($z_{\text{opt}}$)} & \multicolumn{1}{l}{gap (\%)} & GRB & \textbf{r\textbf{ABM}} & \textbf{ABM} & \textbf{\textbf{ABM}-i5} & \textbf{\textbf{\textbf{ABM}-i1}0} & AC & GRD & RND \\
\midrule
Synthetic-100-2 & 100 & 2 & 12,046.92 & -1.57 & 0.00 & 0.76 & 0.01 & 0.00 & 0.00 & 0.06 & 2.03 & 41.87 \\
Synthetic-100-5 & 100 & 5 & 24,473.15 & -0.19 & 0.00 & 1.28 & 0.02 & 0.00 & 0.00 & 0.57 & 7.34 & 35.11 \\
Synthetic-100-10 & 100 & 10 & 52,869.55 & -0.53 & 0.00 & 1.38 & 0.33 & 0.08 & 0.00 & 1.29 & 0.01 & 35.83 \\
Synthetic-1000-2 & 1,000 & 2 & 121,095.14 & -1.91 & 0.00 & 0.05 & 0.00 & 0.00 & 0.00 & 0.12 & 2.23 & 44.97 \\
Synthetic-1000-5 & 1,000 & 5 & 245,144.11 & -0.30 & 0.00 & 0.33 & 0.00 & 0.00 & 0.00 & 0.95 & 7.46 & 38.90 \\
Synthetic-1000-10 & 1,000 & 10 & 529,425.22 & -0.67 & 0.00 & 0.24 & 0.06 & 0.00 & 0.00 & 0.90 & 0.01 & 36.73 \\
Synthetic-2000-2 & 2,000 & 2 & 242,293.64 & -1.89 & 0.00 & 0.03 & 0.00 & 0.00 & 0.00 & 0.09 & 2.24 & 46.78 \\
Synthetic-2000-5 & 2,000 & 5 & 489,931.39 & -0.40 & 0.00 & 0.42 & 0.00 & 0.00 & 0.00 & 0.94 & 7.47 & 38.02 \\
Synthetic-2000-10 & 2,000 & 10 & 1,059,365.74 & -0.55 & 0.00 & 0.27 & 0.05 & 0.00 & 0.00 & 1.00 & 0.01 & 37.60 \\
Synthetic-3000-2 & 3,000 & 2 & 363,489.14 & -1.86 & 0.00 & 0.04 & 0.00 & 0.00 & 0.00 & 0.14 & 2.24 & 45.16 \\
Synthetic-3000-5 & 3,000 & 5 & 735,153.35 & -0.45 & 0.00 & 0.28 & 0.00 & 0.00 & 0.00 & 0.96 & 7.42 & 37.68 \\
Synthetic-3000-10 & 3,000 & 10 & 1,588,665.49 & -0.74 & 0.00 & 0.21 & 0.18 & 0.00 & 0.00 & 0.94 & 0.01 & 37.56 \\
\midrule Average &  &  &  & &  0.00 & 0.44 & 0.05 & 0.01 & 0.00 & 0.66 & 3.20 & 39.68 \\
\bottomrule
\end{tabular}

\scriptsize
\up The approaches {\sf rABM}, \ABM, \ABM-i5, and \ABM-i10 are introduced in this paper. The abbreviations are explained in Table~\ref{tbl:tested-algorithms}. The upper bound (UB) gap is $\frac{z_{\text{opt}} - \textup{UB}}{z_{\text{opt}}} \times 100$.
\end{table}

\begin{table}
\tiny
\caption{Runtimes for synthetic datasets in seconds\label{tbl:cpu_large_synthetic}}
\setlength{\tabcolsep}{12pt}
\begin{tabular}{lrrrrrrrrrr}
\toprule
 & & & \multicolumn{8}{c}{Runtime (s)} \\
\cmidrule(lr){4-11}
Dataset & $n$ & $d$ & GRB & \textbf{r\textbf{ABM}} & \textbf{ABM} & \textbf{\textbf{ABM}-i5} & \textbf{\textbf{\textbf{ABM}-i1}0} & AC & GRD & RND \\
\midrule
Synthetic-100-2 & 100 & 2 & 0.12 & \bfseries$<$0.01&$<$0.01& 0.08 & 0.16 & 0.01 &$<$0.01&$<$0.01\\
Synthetic-100-5 & 100 & 5 & 0.07 &$<$0.01& \bfseries$<$0.01& 0.08 & 0.16 & 0.01 &$<$0.01&$<$0.01\\
Synthetic-100-10 & 100 & 10 & 0.02 &$<$0.01& \bfseries$<$0.01& 0.08 & 0.16 & 0.01 &$<$0.01&$<$0.01\\
Synthetic-1000-2 & 1,000 & 2 & 4.24 & 0.04 & \bfseries 0.03 & 0.19 & 0.34 & 0.64 & 0.38 &$<$0.01\\
Synthetic-1000-5 & 1,000 & 5 & 3.28 & 0.03 & \bfseries 0.03 & 0.18 & 0.34 & 0.98 & 0.39 &$<$0.01\\
Synthetic-1000-10 & 1,000 & 10 & 2.68 & 0.03 & \bfseries 0.02 & 0.18 & 0.33 & 1.11 & 0.38 &$<$0.01\\
Synthetic-2000-2 & 2,000 & 2 & 18.69 & 0.24 & \bfseries 0.19 & 1.16 & 2.11 & 4.95 & 1.75 &$<$0.01\\
Synthetic-2000-5 & 2,000 & 5 & 14.45 & 0.22 & \bfseries 0.19 & 1.10 & 2.03 & 7.64 & 1.73 &$<$0.01\\
Synthetic-2000-10 & 2,000 & 10 & 12.54 & 0.17 & \bfseries 0.12 & 0.78 & 1.44 & 6.70 & 1.72 &$<$0.01\\
Synthetic-3000-2 & 3,000 & 2 & 33.49 & 0.71 & \bfseries 0.62 & 3.51 & 6.38 & 16.57 & 4.18 &$<$0.01\\
Synthetic-3000-5 & 3,000 & 5 & 28.57 & 0.66 & \bfseries 0.59 & 3.40 & 6.26 & 25.74 & 4.24 &$<$0.01\\
Synthetic-3000-10 & 3,000 & 10 & 24.32 & 0.50 & \bfseries 0.45 & 2.36 & 4.31 & 23.92 & 4.18 &$<$0.01\\
\midrule Average &  &  & 11.87 & 0.22 & \bfseries 0.19 & 1.09 & 2.00 & 7.35 & 1.58 &$<$0.01\\
\bottomrule
\end{tabular}

\scriptsize
\up The approaches {\sf rABM}, \ABM, \ABM-i5, and \ABM-i10 are introduced in this paper. The lowest runtime per dataset is shown in bold (excluding RND).
\end{table}

We also repeated the ablation study with the Manhattan ($L_1$) and Euclidean ($L_2$) distances, and the results were qualitatively unchanged, so our conclusions hold for these distances as well (see \url{https://github.com/phil85/abm}). 

\subsection{Comparison to the algorithm of \texorpdfstring{\citet{stanvek2026new}}{Stan\v{e}k and Arustamyan (2026)}}\label{sec:results-tsp}

The algorithm of \citet{stanvek2026new}, referred to here as SA2D, cannot be tested on the datasets used in the experiments described in Sections~\ref{sec:results-small}--\ref{sec:results-ablation}. This is because SA2D is only applicable to datasets with two features and Euclidean distances as edge weights. The datasets used in those experiments are mostly for data with more than two features, and for the metric of squared Euclidean distances. We therefore compare \ABM\ to SA2D in this section, using two collections of datasets with two features and Euclidean distances as edge weights.

The first dataset collection consists of the 49 traveling salesperson datasets with up to $1{,}000$ objects that were used in the experiments of~\citet{stanvek2026new}. We obtained them from the TSPLIB website (see \citealt{reinelt1995tsplib}). The second collection comprises five large-scale datasets from the University of Waterloo TSP website (see \citealt{waterloo_tsp_data}). The website groups its datasets into categories. We selected one dataset from each of the categories National TSPs, VLSI TSPs, World TSP, Mona Lisa, and USA, choosing the largest one whenever a category contains several. We omitted the category Art TSPs because, like {\sf mona-lisa100K}, its instances are point sets that reproduce images and would therefore not add a new type of data. We thus selected the following datasets: {\sf ch71009}, {\sf mona-lisa100K}, {\sf usa115475}, {\sf lrb744710}, and {\sf world}. These datasets range from approximately $71{,}000$ to $1.9$ million objects, each described by two features. The objects represent the locations of cities and towns ({\sf ch71009}, {\sf usa115475}, and {\sf world}), points that reproduce an image when drawn as dots ({\sf mona-lisa100K}), and points from the layout of a large-scale integrated circuit ({\sf lrb744710}). In both collections, for each dataset containing an odd number of objects, we removed the last object from the dataset.

We begin with the results for the first collection. We ran our assignment-based matching algorithm in six variants: without improvement iterations (\ABM) and with one to five improvement iterations (\ABM-i1, \ldots, \ABM-i5). As benchmarks, we used the exact approach GRB and the heuristics SA2D, GRD, and RND. We compare all methods in terms of solution quality, measured by the objective function value, and runtime.

\begin{table}
\scriptsize
\caption{Objective function values for small TSP datasets \label{tbl:ofv_tsp_small}}
\setlength{\tabcolsep}{3.5pt}
\begin{tabular}{lrrrrrrrrrrrrrr}
\toprule
 & & & & & \multicolumn{10}{c}{Deviation from optimal objective (\%)} \\
\cmidrule(lr){6-15}
 & & & \multicolumn{1}{l}{Optimal} & \multicolumn{1}{l}{Upper} & & & & & & & & & & \\
 & & & \multicolumn{1}{l}{objective} & \multicolumn{1}{l}{bound} & & & & & & & & & & \\
Dataset & $n$ & $d$ & \multicolumn{1}{l}{($z_{\text{opt}}$)} & \multicolumn{1}{l}{gap (\%)} & GRB & \textbf{ABM} & \textbf{\textbf{ABM}-i1} & \textbf{\textbf{ABM}-i2} & \textbf{\textbf{ABM}-i3} & \textbf{\textbf{ABM}-i4} & \textbf{\textbf{ABM}-i5} & SA2D & GRD & RND \\
\midrule
eil51 & 50 & 2 & 1,172.70 & -0.06 & 0.00 & 0.61 & 0.12 & 0.04 & 0.01 & 0.00 & 0.00 & 0.01 & 1.38 & 30.09 \\
berlin52 & 52 & 2 & 19,866.74 & -0.10 & 0.00 & 0.37 & 0.06 & 0.03 & 0.02 & 0.02 & 0.00 & 0.03 & 3.97 & 23.68 \\
st70 & 70 & 2 & 2,675.04 & -0.23 & 0.00 & 0.69 & 0.16 & 0.04 & 0.01 & 0.00 & 0.00 & 0.00 & 1.53 & 28.05 \\
eil76 & 76 & 2 & 1,800.17 & -0.06 & 0.00 & 0.41 & 0.08 & 0.03 & 0.01 & 0.01 & 0.00 & 0.06 & 1.37 & 28.91 \\
pr76 & 76 & 2 & 408,093.14 & -0.05 & 0.00 & 0.51 & 0.10 & 0.03 & 0.01 & 0.01 & 0.00 & 0.00 & 1.46 & 30.43 \\
rat99 & 98 & 2 & 6,051.92 & -0.04 & 0.00 & 0.15 & 0.03 & 0.01 & 0.00 & 0.00 & 0.00 & 0.01 & 0.70 & 34.72 \\
eil101 & 100 & 2 & 2,481.04 & -0.11 & 0.00 & 0.21 & 0.04 & 0.01 & 0.00 & 0.00 & 0.00 & 0.01 & 2.21 & 31.62 \\
kroA100 & 100 & 2 & 126,689.43 & -0.10 & 0.00 & 0.24 & 0.08 & 0.04 & 0.02 & 0.01 & 0.01 & 0.02 & 1.43 & 38.49 \\
kroB100 & 100 & 2 & 123,587.97 & -0.23 & 0.00 & 0.11 & 0.03 & 0.01 & 0.01 & 0.00 & 0.00 & 0.00 & 1.15 & 31.32 \\
kroC100 & 100 & 2 & 127,390.87 & -0.08 & 0.00 & 0.19 & 0.02 & 0.01 & 0.00 & 0.00 & 0.00 & 0.66 & 1.40 & 33.84 \\
kroD100 & 100 & 2 & 118,753.17 & -0.20 & 0.00 & 0.48 & 0.11 & 0.04 & 0.02 & 0.01 & 0.01 & 0.14 & 1.24 & 34.06 \\
kroE100 & 100 & 2 & 128,247.36 & -0.20 & 0.00 & 0.05 & 0.02 & 0.01 & 0.00 & 0.00 & 0.00 & 0.05 & 1.05 & 33.04 \\
rd100 & 100 & 2 & 41,004.97 & -0.14 & 0.00 & 0.25 & 0.06 & 0.02 & 0.01 & 0.00 & 0.00 & 0.00 & 2.11 & 31.63 \\
lin105 & 104 & 2 & 89,163.85 & -0.22 & 0.00 & 0.28 & 0.07 & 0.04 & 0.02 & 0.02 & 0.01 & 0.10 & 0.94 & 31.86 \\
pr107 & 106 & 2 & 458,092.49 & -0.62 & 0.00 & 0.16 & 0.05 & 0.01 & 0.01 & 0.00 & 0.00 & 0.00 & 0.95 & 33.25 \\
pr124 & 124 & 2 & 517,739.08 & -0.44 & 0.00 & 0.68 & 0.15 & 0.07 & 0.01 & 0.00 & 0.00 & 0.28 & 3.65 & 31.43 \\
bier127 & 126 & 2 & 413,148.65 & -0.09 & 0.00 & 0.12 & 0.03 & 0.01 & 0.01 & 0.00 & 0.00 & 0.07 & 2.09 & 27.80 \\
ch130 & 130 & 2 & 33,739.93 & -0.06 & 0.00 & 0.20 & 0.06 & 0.02 & 0.01 & 0.01 & 0.00 & 0.06 & 1.82 & 30.79 \\
pr136 & 136 & 2 & 623,472.80 & -0.01 & 0.00 & 0.53 & 0.13 & 0.03 & 0.02 & 0.01 & 0.01 & 0.00 & 0.92 & 36.94 \\
pr144 & 144 & 2 & 600,981.27 & -0.08 & 0.00 & 0.10 & 0.02 & 0.01 & 0.00 & 0.00 & 0.00 & 0.00 & 1.76 & 31.31 \\
ch150 & 150 & 2 & 39,283.09 & -0.08 & 0.00 & 0.10 & 0.02 & 0.01 & 0.00 & 0.00 & 0.00 & 0.00 & 0.65 & 33.05 \\
kroA150 & 150 & 2 & 191,046.67 & -0.04 & 0.00 & 0.20 & 0.06 & 0.01 & 0.01 & 0.00 & 0.00 & 0.01 & 0.93 & 33.11 \\
kroB150 & 150 & 2 & 192,690.89 & -0.10 & 0.00 & 0.15 & 0.02 & 0.01 & 0.00 & 0.00 & 0.00 & 0.07 & 1.18 & 33.44 \\
pr152 & 152 & 2 & 747,963.34 & -0.03 & 0.00 & 0.10 & 0.02 & 0.00 & 0.00 & 0.00 & 0.00 & 0.03 & 1.15 & 32.40 \\
u159 & 158 & 2 & 330,587.12 & -0.11 & 0.00 & 0.29 & 0.05 & 0.02 & 0.00 & 0.00 & 0.00 & 0.15 & 1.28 & 30.59 \\
rat195 & 194 & 2 & 16,568.54 & -0.01 & 0.00 & 0.15 & 0.02 & 0.00 & 0.00 & 0.00 & 0.00 & 0.00 & 0.43 & 32.80 \\
d198 & 198 & 2 & 129,853.84 & -0.19 & 0.00 & 0.03 & 0.01 & 0.00 & 0.00 & 0.00 & 0.00 & 0.06 & 1.79 & 25.87 \\
kroA200 & 200 & 2 & 254,477.47 & -0.06 & 0.00 & 0.37 & 0.05 & 0.02 & 0.01 & 0.00 & 0.00 & 0.16 & 0.70 & 34.72 \\
kroB200 & 200 & 2 & 245,569.48 & -0.13 & 0.00 & 0.10 & 0.01 & 0.00 & 0.00 & 0.00 & 0.00 & 0.01 & 0.76 & 34.44 \\
ts225 & 224 & 2 & 1,155,655.73 & 0.00 & 0.00 & 0.00 & 0.00 & 0.00 & 0.00 & 0.00 & 0.00 & 0.01 & 0.04 & 29.95 \\
tsp225 & 224 & 2 & 29,623.54 & -0.21 & 0.00 & 0.02 & 0.01 & 0.00 & 0.00 & 0.00 & 0.00 & 0.08 & 0.77 & 31.78 \\
pr226 & 226 & 2 & 1,280,860.79 & -0.21 & 0.00 & 0.17 & 0.05 & 0.02 & 0.02 & 0.01 & 0.01 & 0.00 & 1.52 & 30.24 \\
gil262 & 262 & 2 & 19,587.54 & -0.03 & 0.00 & 0.10 & 0.02 & 0.01 & 0.00 & 0.00 & 0.00 & 0.00 & 1.05 & 32.36 \\
pr264 & 264 & 2 & 909,995.74 & -0.30 & 0.00 & 0.10 & 0.01 & 0.00 & 0.00 & 0.00 & 0.00 & 0.00 & 0.39 & 40.26 \\
a280 & 280 & 2 & 25,323.58 & -0.10 & 0.00 & 0.04 & 0.01 & 0.00 & 0.00 & 0.00 & 0.00 & 0.00 & 0.58 & 30.80 \\
pr299 & 298 & 2 & 564,785.49 & -0.19 & 0.00 & 0.06 & 0.01 & 0.00 & 0.00 & 0.00 & 0.00 & 0.02 & 0.85 & 33.90 \\
lin318 & 318 & 2 & 430,249.73 & -0.06 & 0.00 & 0.08 & 0.02 & 0.01 & 0.00 & 0.00 & 0.00 & 0.01 & 0.87 & 32.70 \\
rd400 & 400 & 2 & 155,832.84 & -0.04 & 0.00 & 0.09 & 0.01 & 0.00 & 0.00 & 0.00 & 0.00 & 0.00 & 0.61 & 30.06 \\
fl417 & 416 & 2 & 388,512.83 & -0.16 & 0.00 & 0.06 & 0.01 & 0.00 & 0.00 & 0.00 & 0.00 & 1.31 & 0.49 & 36.67 \\
pr439 & 438 & 2 & 1,352,452.87 & -0.13 & 0.00 & 0.07 & 0.01 & 0.00 & 0.00 & 0.00 & 0.00 & 0.03 & 3.78 & 31.10 \\
pcb442 & 442 & 2 & 564,807.22 & -0.03 & 0.00 & 0.07 & 0.01 & 0.00 & 0.00 & 0.00 & 0.00 & 0.00 & 0.58 & 31.22 \\
d493 & 492 & 2 & 314,930.52 & -0.03 & 0.00 & 0.02 & 0.00 & 0.00 & 0.00 & 0.00 & 0.00 & 0.01 & 3.87 & 29.53 \\
rat575 & 574 & 2 & 83,537.75 & 0.00 & 0.00 & 0.07 & 0.01 & 0.00 & 0.00 & 0.00 & 0.00 & 0.00 & 0.25 & 32.55 \\
u574 & 574 & 2 & 498,492.45 & -0.03 & 0.00 & 0.10 & 0.01 & 0.00 & 0.00 & 0.00 & 0.00 & 0.12 & 0.42 & 32.34 \\
p654 & 654 & 2 & 1,630,061.99 & -0.07 & 0.00 & 0.04 & 0.00 & 0.00 & 0.00 & 0.00 & 0.00 & 0.01 & 0.18 & 38.93 \\
d657 & 656 & 2 & 623,285.26 & -0.06 & 0.00 & 0.02 & 0.00 & 0.00 & 0.00 & 0.00 & 0.00 & 0.00 & 0.89 & 31.16 \\
u724 & 724 & 2 & 642,726.05 & -0.01 & 0.00 & 0.02 & 0.00 & 0.00 & 0.00 & 0.00 & 0.00 & 0.00 & 0.39 & 31.70 \\
rat783 & 782 & 2 & 131,936.63 & 0.00 & 0.00 & 0.02 & 0.00 & 0.00 & 0.00 & 0.00 & 0.00 & 0.00 & 0.22 & 32.79 \\
dsj1000 & 1,000 & 2 & 403,067,454.68 & -0.98 & 0.00 & 0.01 & 0.00 & 0.00 & 0.00 & 0.00 & 0.00 & 1.00 & 2.81 & 29.72 \\
\midrule Average &  &  &  & &  0.00 & 0.18 & 0.04 & 0.01 & 0.01 & 0.00 & 0.00 & 0.09 & 1.28 & 32.11 \\
\bottomrule
\end{tabular}

\up \ABM\ and \ABM-iX are introduced in this paper. The abbreviations are explained in Table~\ref{tbl:tested-algorithms}. The upper bound (UB) gap is $\frac{z_{\text{opt}} - \textup{UB}}{z_{\text{opt}}} \times 100$.
\end{table}

Table~\ref{tbl:ofv_tsp_small} reports the solution quality. The first three columns give, for each dataset, its name, the number of objects ($n$), and the number of features ($d$). The fourth column reports the optimal objective function value for each instance. The fifth column reports the relative gap between the upper bound (UB) and the optimal objective function value $z_{\text{opt}}$, computed as $\frac{z_{\text{opt}} - \textup{UB}}{z_{\text{opt}}} \times 100$. The remaining columns report, for each approach, the percentage deviation from this optimal value, computed as $\%DEV = \frac{z_{\text{opt}} - z_{\text{approach}}}{z_{\text{opt}}} \times 100$. The last row states the average deviation for each approach. The upper bound gaps to the known optima are very small (0.13\% on average, at most 0.98\%), showing that the bound is tight on these instances.

The exact approach GRB finds certified optimal solutions for all datasets in less than two seconds. Without improvement iterations, \ABM\ achieves an average deviation of $0.18\%$ from the optimum. A single improvement iteration reduces this average to $0.04\%$ (\ABM-i1), and with five improvement iterations the deviation is at most $0.01\%$ on every dataset (\ABM-i5). SA2D attains a lower average deviation than \ABM\ ($0.09\%$), but its deviations vary more: it deviates by $1.31\%$ on {\sf fl417}, $1.00\%$ on {\sf dsj1000}, and $0.66\%$ on {\sf kroC100}, whereas the largest deviation of \ABM\ is $0.69\%$ and that of \ABM-i1 is $0.16\%$. The deviations of SA2D agree with those reported by~\citet{stanvek2026new} for the same datasets. The deviations of \ABM\ are largest for the smallest datasets and decrease with $n$, whereas the largest deviations of SA2D occur on structured datasets such as {\sf fl417}. With at least one improvement iteration, the assignment-based approaches thus achieve both a lower average and a lower maximum deviation than SA2D. As in the previous experiments, GRD performs slightly worse than the best heuristics, with an average deviation of $1.28\%$, whereas random matching (RND) performs considerably worse, with an average deviation of $32.11\%$.

Table~\ref{tbl:cpu_tsp_small} reports the runtimes in seconds for the small TSP datasets. All approaches solve every dataset quickly: the heuristics require at most $0.27$ seconds and GRB at most $1.64$ seconds. The runtime of GRB grows noticeably with the number of objects, whereas that of \ABM\ stays below $0.05$ seconds for all datasets. RND is the fastest approach, followed by \ABM, SA2D, and GRD, although the runtime differences among these four methods are small on these instances. Each improvement iteration increases the average runtime of \ABM\ by about $0.02$ seconds.

\begin{table}
\scriptsize
\caption{Runtimes for small TSP datasets in seconds\label{tbl:cpu_tsp_small}}
\setlength{\tabcolsep}{7pt}
\begin{tabular}{lrrrrrrrrrrrr}
\toprule
 & & & \multicolumn{10}{c}{Runtime (s)} \\
\cmidrule(lr){4-13}
Dataset & $n$ & $d$ & GRB & \textbf{ABM} & \textbf{\textbf{ABM}-i1} & \textbf{\textbf{ABM}-i2} & \textbf{\textbf{ABM}-i3} & \textbf{\textbf{ABM}-i4} & \textbf{\textbf{ABM}-i5} & SA2D & GRD & RND \\
\midrule
eil51 & 50 & 2 &$<$0.01& \bfseries$<$0.01& 0.02 & 0.03 & 0.05 & 0.07 & 0.08 &$<$0.01&$<$0.01&$<$0.01\\
berlin52 & 52 & 2 &$<$0.01& \bfseries$<$0.01& 0.02 & 0.03 & 0.05 & 0.07 & 0.08 &$<$0.01&$<$0.01&$<$0.01\\
st70 & 70 & 2 & 0.01 & \bfseries$<$0.01& 0.02 & 0.03 & 0.05 & 0.07 & 0.08 &$<$0.01&$<$0.01&$<$0.01\\
eil76 & 76 & 2 & 0.01 & \bfseries$<$0.01& 0.02 & 0.04 & 0.05 & 0.07 & 0.08 &$<$0.01&$<$0.01&$<$0.01\\
pr76 & 76 & 2 & 0.01 & \bfseries$<$0.01& 0.02 & 0.03 & 0.05 & 0.07 & 0.08 &$<$0.01&$<$0.01&$<$0.01\\
rat99 & 98 & 2 & 0.01 & \bfseries$<$0.01& 0.02 & 0.03 & 0.05 & 0.07 & 0.08 &$<$0.01&$<$0.01&$<$0.01\\
eil101 & 100 & 2 & 0.01 & \bfseries$<$0.01& 0.02 & 0.03 & 0.05 & 0.07 & 0.08 &$<$0.01&$<$0.01&$<$0.01\\
kroA100 & 100 & 2 & 0.01 & \bfseries$<$0.01& 0.02 & 0.03 & 0.05 & 0.07 & 0.08 &$<$0.01&$<$0.01&$<$0.01\\
kroB100 & 100 & 2 & 0.01 & \bfseries$<$0.01& 0.02 & 0.04 & 0.05 & 0.07 & 0.08 &$<$0.01&$<$0.01&$<$0.01\\
kroC100 & 100 & 2 & 0.01 & \bfseries$<$0.01& 0.02 & 0.04 & 0.05 & 0.07 & 0.08 &$<$0.01&$<$0.01&$<$0.01\\
kroD100 & 100 & 2 & 0.01 & \bfseries$<$0.01& 0.02 & 0.03 & 0.05 & 0.07 & 0.08 &$<$0.01&$<$0.01&$<$0.01\\
kroE100 & 100 & 2 & 0.01 & \bfseries$<$0.01& 0.02 & 0.03 & 0.05 & 0.07 & 0.08 &$<$0.01&$<$0.01&$<$0.01\\
rd100 & 100 & 2 & 0.01 & \bfseries$<$0.01& 0.02 & 0.04 & 0.05 & 0.07 & 0.08 &$<$0.01&$<$0.01&$<$0.01\\
lin105 & 104 & 2 & 0.01 & \bfseries$<$0.01& 0.02 & 0.03 & 0.05 & 0.07 & 0.08 &$<$0.01&$<$0.01&$<$0.01\\
pr107 & 106 & 2 & 0.01 & \bfseries$<$0.01& 0.02 & 0.03 & 0.05 & 0.07 & 0.08 &$<$0.01&$<$0.01&$<$0.01\\
pr124 & 124 & 2 & 0.02 & \bfseries$<$0.01& 0.02 & 0.04 & 0.05 & 0.07 & 0.08 &$<$0.01&$<$0.01&$<$0.01\\
bier127 & 126 & 2 & 0.02 & \bfseries$<$0.01& 0.02 & 0.03 & 0.05 & 0.07 & 0.08 &$<$0.01&$<$0.01&$<$0.01\\
ch130 & 130 & 2 & 0.02 & \bfseries$<$0.01& 0.02 & 0.04 & 0.05 & 0.07 & 0.08 &$<$0.01& 0.01 &$<$0.01\\
pr136 & 136 & 2 & 0.02 & \bfseries$<$0.01& 0.02 & 0.03 & 0.05 & 0.07 & 0.08 &$<$0.01& 0.01 &$<$0.01\\
pr144 & 144 & 2 & 0.03 & \bfseries$<$0.01& 0.02 & 0.03 & 0.05 & 0.07 & 0.08 &$<$0.01& 0.01 &$<$0.01\\
ch150 & 150 & 2 & 0.03 & \bfseries$<$0.01& 0.02 & 0.03 & 0.05 & 0.07 & 0.08 &$<$0.01& 0.01 &$<$0.01\\
kroA150 & 150 & 2 & 0.03 & \bfseries$<$0.01& 0.02 & 0.03 & 0.05 & 0.07 & 0.08 &$<$0.01& 0.01 &$<$0.01\\
kroB150 & 150 & 2 & 0.02 & \bfseries$<$0.01& 0.02 & 0.04 & 0.05 & 0.07 & 0.08 &$<$0.01& 0.01 &$<$0.01\\
pr152 & 152 & 2 & 0.03 & \bfseries$<$0.01& 0.02 & 0.03 & 0.05 & 0.07 & 0.08 &$<$0.01& 0.01 &$<$0.01\\
u159 & 158 & 2 & 0.03 & \bfseries$<$0.01& 0.02 & 0.03 & 0.05 & 0.07 & 0.08 &$<$0.01& 0.01 &$<$0.01\\
rat195 & 194 & 2 & 0.06 & \bfseries$<$0.01& 0.02 & 0.03 & 0.05 & 0.07 & 0.08 & 0.01 & 0.01 &$<$0.01\\
d198 & 198 & 2 & 0.06 & \bfseries$<$0.01& 0.02 & 0.03 & 0.05 & 0.07 & 0.08 & 0.01 & 0.01 &$<$0.01\\
kroA200 & 200 & 2 & 0.04 & \bfseries$<$0.01& 0.02 & 0.04 & 0.05 & 0.07 & 0.08 & 0.01 & 0.01 &$<$0.01\\
kroB200 & 200 & 2 & 0.05 & \bfseries$<$0.01& 0.02 & 0.03 & 0.05 & 0.07 & 0.08 & 0.01 & 0.01 &$<$0.01\\
ts225 & 224 & 2 & 0.06 & \bfseries$<$0.01& 0.02 & 0.03 & 0.05 & 0.07 & 0.08 & 0.01 & 0.01 &$<$0.01\\
tsp225 & 224 & 2 & 0.07 & \bfseries$<$0.01& 0.02 & 0.03 & 0.05 & 0.07 & 0.08 & 0.01 & 0.01 &$<$0.01\\
pr226 & 226 & 2 & 0.06 & \bfseries$<$0.01& 0.02 & 0.03 & 0.05 & 0.07 & 0.08 & 0.01 & 0.01 &$<$0.01\\
gil262 & 262 & 2 & 0.11 & \bfseries$<$0.01& 0.02 & 0.03 & 0.05 & 0.07 & 0.08 & 0.01 & 0.02 &$<$0.01\\
pr264 & 264 & 2 & 0.11 & \bfseries$<$0.01& 0.02 & 0.03 & 0.05 & 0.07 & 0.08 & 0.01 & 0.02 &$<$0.01\\
a280 & 280 & 2 & 0.14 & \bfseries$<$0.01& 0.02 & 0.03 & 0.05 & 0.07 & 0.08 & 0.01 & 0.02 &$<$0.01\\
pr299 & 298 & 2 & 0.19 & \bfseries$<$0.01& 0.02 & 0.04 & 0.05 & 0.07 & 0.08 & 0.01 & 0.03 &$<$0.01\\
lin318 & 318 & 2 & 0.17 & \bfseries$<$0.01& 0.02 & 0.04 & 0.05 & 0.07 & 0.08 & 0.01 & 0.03 &$<$0.01\\
rd400 & 400 & 2 & 0.28 & \bfseries$<$0.01& 0.02 & 0.04 & 0.05 & 0.07 & 0.09 & 0.01 & 0.05 &$<$0.01\\
fl417 & 416 & 2 & 0.31 & \bfseries$<$0.01& 0.02 & 0.04 & 0.05 & 0.07 & 0.09 & 0.01 & 0.05 &$<$0.01\\
pr439 & 438 & 2 & 0.46 & \bfseries$<$0.01& 0.02 & 0.04 & 0.05 & 0.07 & 0.09 & 0.01 & 0.06 &$<$0.01\\
pcb442 & 442 & 2 & 0.32 & \bfseries$<$0.01& 0.02 & 0.04 & 0.05 & 0.07 & 0.09 & 0.01 & 0.06 &$<$0.01\\
d493 & 492 & 2 & 0.53 & \bfseries 0.01 & 0.02 & 0.04 & 0.06 & 0.07 & 0.09 & 0.02 & 0.08 &$<$0.01\\
rat575 & 574 & 2 & 0.83 & \bfseries 0.01 & 0.03 & 0.04 & 0.06 & 0.08 & 0.09 & 0.02 & 0.11 &$<$0.01\\
u574 & 574 & 2 & 0.70 & \bfseries 0.01 & 0.03 & 0.04 & 0.06 & 0.08 & 0.09 & 0.02 & 0.11 &$<$0.01\\
p654 & 654 & 2 & 0.89 & \bfseries 0.01 & 0.03 & 0.04 & 0.06 & 0.07 & 0.09 & 0.02 & 0.13 &$<$0.01\\
d657 & 656 & 2 & 0.89 & \bfseries 0.01 & 0.03 & 0.05 & 0.06 & 0.08 & 0.09 & 0.02 & 0.14 &$<$0.01\\
u724 & 724 & 2 & 1.40 & \bfseries 0.02 & 0.05 & 0.08 & 0.11 & 0.14 & 0.17 & 0.02 & 0.18 &$<$0.01\\
rat783 & 782 & 2 & 1.64 & \bfseries 0.02 & 0.05 & 0.08 & 0.11 & 0.15 & 0.18 & 0.02 & 0.21 &$<$0.01\\
dsj1000 & 1,000 & 2 & 1.43 & \bfseries 0.04 & 0.09 & 0.14 & 0.18 & 0.23 & 0.27 & 0.04 & 0.39 &$<$0.01\\
\midrule Average &  &  & 0.23 & \bfseries$<$0.01& 0.02 & 0.04 & 0.06 & 0.07 & 0.09 & 0.01 & 0.04 &$<$0.01\\
\bottomrule
\end{tabular}

\up \ABM\ and \ABM-iX are introduced in this paper. The lowest runtime per dataset is shown in bold (excluding RND).
\end{table}

Next, we report the results for the second collection, which comprises the large-scale TSP datasets. We ran \ABM\ with assignment problem sizes of $1{,}000$ (\ABM-d1000) and $2{,}000$ (\ABM-d2000). Since the exact approach GRB did not find a feasible solution for any of these datasets within a time limit of $7{,}200$ seconds, we used the exact approach HEX-L instead, with the same time limit. As heuristic benchmarks, we used SA2D and RND. GRD exceeded the available memory on all large-scale instances and is therefore omitted from the tables. As before, we compare the tested methods in terms of solution quality and runtime. 

Table~\ref{tbl:ofv_tsp_large} reports the solution quality. It is structured analogously to the corresponding table for the small TSP datasets. Since HEX-L reached the time limit on all datasets, the best objective values are not proven optimal. \ABM-d2000 achieves the best average solution quality, followed by \ABM-d1000 and SA2D. All three stay within 0.11\% of the best objective value on every dataset. HEX-L deviates by less than 3\% on the three smaller datasets but by more than 23\% on {\sf lrb744710} and {\sf world}, which is close to the deviation of RND. With upper bound gaps of at most 0.29\% (0.21\% on average), the best solutions are provably near-optimal for all five instances.

Table~\ref{tbl:cpu_tsp_large} reports the runtimes in seconds. Since the subproblems of \ABM-d1000 and \ABM-d2000 are independent, we solve them in parallel on 16 cores, whereas SA2D runs on a single thread. The reported runtimes are therefore not directly comparable. With parallelization, the runtime advantage of \ABM-d1000 over SA2D grows with the size of the dataset: on {\sf world}, \ABM-d1000 requires 23.27 seconds, whereas SA2D requires 160.87 seconds. The runtimes of \ABM-d1000 and \ABM-d2000 grow roughly linearly with the number of objects, whereas that of SA2D appears to grow at a rate consistent with its $\mathcal{O}(n \log n)$ complexity.

Overall, the assignment-based approaches compare favorably with SA2D. On the small TSP datasets, \ABM\ with at least one improvement iteration achieves both a lower average and a lower maximum deviation than SA2D at a comparable runtime, within a few hundredths of a second. On the large-scale TSP datasets, \ABM-d1000 and \ABM-d2000 achieve a better average solution quality than SA2D and, with the subproblems solved in parallel, shorter runtimes, with the advantage growing with the size of the dataset. These results are notable because SA2D is designed specifically for maximum weight matching in the Euclidean plane, whereas \ABM\ and its variants handle any number of features and any norm-induced distance.

\begin{table}
\scriptsize
\caption{Objective function values for large-scale TSP datasets \label{tbl:ofv_tsp_large}}
\setlength{\tabcolsep}{8.4pt}
\begin{tabular}{lrrrrrrrrr}
\toprule
 & & & & & \multicolumn{5}{c}{Deviation from best objective (\%)} \\
\cmidrule(lr){6-10}
 & & & \multicolumn{1}{l}{Best} & \multicolumn{1}{l}{Upper} & & & & & \\
 & & & \multicolumn{1}{l}{objective} & \multicolumn{1}{l}{bound} & & & & & \\
Dataset & $n$ & $d$ & \multicolumn{1}{l}{($z_{\text{best}}$)} & \multicolumn{1}{l}{gap (\%)} & HEX-L & \textbf{\textbf{ABM}-d1000} & \textbf{\textbf{ABM}-d2000} & SA2D & RND \\
\midrule
ch71009 & 71,008 & 2 & 569,449,557.34 & -0.09 & 1.90 & 0.04 & 0.01 & 0.00 & 28.57 \\
mona-lisa100K & 100,000 & 2 & 725,240,776.62 & -0.29 & 1.58 & 0.03 & 0.00 & 0.03 & 31.74 \\
usa115475 & 115,474 & 2 & 1,286,895,045.46 & -0.20 & 2.84 & 0.02 & 0.00 & 0.11 & 28.24 \\
lrb744710 & 744,710 & 2 & 927,248,678.16 & -0.25 & 26.84 & 0.03 & 0.00 & 0.00 & 32.20 \\
world & 1,904,710 & 2 & 108,486,003.57 & -0.23 & 23.56 & 0.02 & 0.00 & 0.09 & 28.34 \\
\midrule Average &  &  &  & &  11.35 & 0.03 & 0.00 & 0.05 & 29.82 \\
\bottomrule
\end{tabular}

\up \ABM-d1000 and \ABM-d2000 are introduced in this paper. The abbreviations are explained in Table~\ref{tbl:tested-algorithms}. The upper bound (UB) gap is $\frac{z_{\text{best}} - \textup{UB}}{z_{\text{best}}} \times 100$.
\end{table}

\begin{table}
\scriptsize
\caption{Runtimes for large-scale TSP datasets in seconds\label{tbl:cpu_tsp_large}}
\setlength{\tabcolsep}{15.2pt}
\begin{tabular}{lrrrrrrr}
\toprule
 & & & \multicolumn{5}{c}{Runtime (s)} \\
\cmidrule(lr){4-8}
Dataset & $n$ & $d$ & HEX-L & \textbf{\textbf{ABM}-d1000} & \textbf{\textbf{ABM}-d2000} & SA2D & RND \\
\midrule
ch71009 & 71,008 & 2 & 7,201.30 & \bfseries 1.05 & 4.87 & 4.57 & 0.01 \\
mona-lisa100K & 100,000 & 2 & 7,202.07 & \bfseries 1.31 & 5.07 & 6.92 & 0.01 \\
usa115475 & 115,474 & 2 & 7,202.00 & \bfseries 1.47 & 5.82 & 7.75 & 0.01 \\
lrb744710 & 744,710 & 2 & 7,217.21 & \bfseries 8.32 & 32.28 & 62.90 & 0.06 \\
world & 1,904,710 & 2 & 7,251.79 & \bfseries 23.27 & 93.81 & 160.87 & 0.14 \\
\midrule Average &  &  & 7,214.87 & \bfseries 7.09 & 28.37 & 48.60 & 0.04 \\
\bottomrule
\end{tabular}

\up The \ABM-dX approaches are introduced in this paper. The lowest runtime per dataset is shown in bold (excluding RND).
\end{table}

\section{Conclusions}\label{sec:conclusions}
We present here a new heuristic, assignment-based matching (\ABM), for the maximum weight non-bipartite matching problem on Euclidean graphs. \ABM\ obtains a solution by solving a single weighted bipartite matching (assignment) problem, which can then be improved by solving additional assignment problems. 
For large-scale instances with more than $100{,}000$ objects, we present a decomposition strategy that drastically improves the scalability of \ABM. An extensive empirical study shows that \ABM\ produces near-optimal solutions and is considerably faster than all existing exact and heuristic methods. Upper bounds on the optimal objective value certify the quality of these solutions where the optimum is unknown: the gap to the bound is at most 7\% and typically much smaller. Consequently, \ABM\ can solve the Euclidean maximum weight non-bipartite matching problem for datasets far larger than was previously possible, making it a powerful tool for large-scale geometric applications.

Several directions for future research remain. One is to investigate other strategies of splitting the objects into the two bipartite sets. For example, a hyperplane through the centroid, orthogonal to the first principal component, would place the objects on either side of the centroid in different sets. Another direction is to combine the decomposition with the batched improvement procedure, which could further improve solutions on the largest instances.


\bibliographystyle{elsarticle-harv} 
\bibliography{literature}

\section*{Appendix: proof of Proposition~\ref{prop:half}}
This appendix proves Proposition~\ref{prop:half}, which states that \ABM\ is a $1/2$-approximation for MWM with squared Euclidean distances. The proof compares both the optimal matching and the matching returned by \ABM\ with the total scatter $\mathrm{SSE}$ of the data: Lemma~\ref{lem:upper} shows that $\mathrm{OPT} \le 2\,\mathrm{SSE}$, and Lemma~\ref{lem:lower} 
shows that the matching solution delivered by \ABM\ is of weight at least $\mathrm{SSE}$.

\paragraph{Notation} Let $\bm{x}_1, \dots, \bm{x}_n \in \mathbb{R}^d$ with $n$ even, $\bm{\mu} = \frac{1}{n} \sum_{i} \bm{x}_i$, and $\mathrm{SSE} = \sum_{i} \|\bm{x}_i - \bm{\mu}\|^2$. For a perfect matching $M$, let $f(M) = \sum_{\{i,j\} \in M} \|\bm{x}_i - \bm{x}_j\|^2$, and let
$\mathrm{OPT} = \max_M f(M)$. For a set $S$ of objects, $\bm{\mu}_S$ and $\mathrm{SSE}_S = \sum_{i \in S} \|\bm{x}_i - \bm{\mu}_S\|^2$ denote its centroid and its scatter around that centroid.
 
\begin{lemma}[Upper bound]\label{lem:upper}
Every perfect matching $M$ satisfies $f(M) \le 2\,\mathrm{SSE}$. In particular, $\mathrm{OPT} \le 2\,\mathrm{SSE}$.
\end{lemma}
 
\begin{proof}
For any vectors $\bm{a}, \bm{b}$, the parallelogram law gives $\|\bm{a} - \bm{b}\|^2 + \|\bm{a} + \bm{b}\|^2 = 2\|\bm{a}\|^2 + 2\|\bm{b}\|^2$, hence $\|\bm{a} - \bm{b}\|^2 \le 2\|\bm{a}\|^2 + 2\|\bm{b}\|^2$. With $\bm{a} = \bm{x}_i - \bm{\mu}$ and $\bm{b} = \bm{x}_j - \bm{\mu}$, we obtain $\|\bm{x}_i - \bm{x}_j\|^2 \le 2\|\bm{x}_i - \bm{\mu}\|^2 + 2\|\bm{x}_j - \bm{\mu}\|^2$. Summing over the pairs of $M$, in which every object appears exactly once, yields $f(M) \le 2\,\mathrm{SSE}$.
\end{proof}
 
\begin{lemma}[Lower bound]\label{lem:lower}
Let $A, B$ be a partition of the objects with $|A| = |B| = m = n/2$, and let $M^\ast$ be a maximum-weight perfect matching between $A$ and $B$. Then
\[
  f(M^\ast) \;\ge\; \mathrm{SSE} + \tfrac{m}{2}\,\|\bm{\mu}_A - \bm{\mu}_B\|^2
  \;\ge\; \mathrm{SSE}.
\]
\end{lemma}
 
\begin{proof}
Each pair $(\bm{a}, \bm{b}) \in A \times B$ is contained in exactly $(m-1)!$ of the $m!$ perfect matchings between $A$ and $B$. Their average objective value is therefore $\frac{1}{m} \sum_{\bm{a} \in A} \sum_{\bm{b} \in B} \|\bm{a} - \bm{b}\|^2$,
and $f(M^\ast)$ is at least this average. For fixed $\bm{a}$, since $\sum_{\bm{b} \in B} (\bm{b} - \bm{\mu}_B) = \bm{0}$, the cross term vanishes in
\[
  \sum_{\bm{b} \in B} \|\bm{a} - \bm{b}\|^2
  = \sum_{\bm{b} \in B} \|(\bm{a} - \bm{\mu}_B) - (\bm{b} - \bm{\mu}_B)\|^2
  = m\,\|\bm{a} - \bm{\mu}_B\|^2 + \mathrm{SSE}_B .
\]
By the same argument, $\sum_{\bm{a} \in A} \|\bm{a} - \bm{\mu}_B\|^2 = \mathrm{SSE}_A + m\,\|\bm{\mu}_A - \bm{\mu}_B\|^2$.
Hence
\[
  \frac{1}{m} \sum_{\bm{a} \in A} \sum_{\bm{b} \in B} \|\bm{a} - \bm{b}\|^2
  = \mathrm{SSE}_A + \mathrm{SSE}_B + m\,\|\bm{\mu}_A - \bm{\mu}_B\|^2 .
\]
The within/between decomposition of the total scatter gives $\mathrm{SSE} = \mathrm{SSE}_A + \mathrm{SSE}_B + m\,\|\bm{\mu}_A - \bm{\mu}\|^2 + m\,\|\bm{\mu}_B - \bm{\mu}\|^2$. Since $|A| = |B|$, we have $\bm{\mu} = (\bm{\mu}_A + \bm{\mu}_B)/2$, so each
between term equals $\frac{m}{4}\|\bm{\mu}_A - \bm{\mu}_B\|^2$. Therefore the average equals $\mathrm{SSE} + \frac{m}{2}\|\bm{\mu}_A - \bm{\mu}_B\|^2$.
\end{proof}
 
\begin{proof}[Proof of Proposition~\ref{prop:half}]
\ABM\ returns $M^\ast$ for its partition, which is a perfect matching of all $n$ objects. By Lemmas~\ref{lem:lower} and~\ref{lem:upper}, $f(M^\ast) \ge \mathrm{SSE} \ge \mathrm{OPT}/2$. Since Lemma~\ref{lem:lower} holds for every partition into two sets of equal size, the bound does not depend on how \ABM\ chooses its partition.
\end{proof}
 
\paragraph{Tightness} For small $\varepsilon > 0$ (e.g., $\varepsilon \le 0.05$), consider $\bm{x}_1 = (1, -2\varepsilon)$, $\bm{x}_2 = (0, 1-\varepsilon)$, $\bm{x}_3 = (-1+2\varepsilon, -2\varepsilon)$, and $\bm{x}_4 = (-2\varepsilon, -1+5\varepsilon)$. Their centroid is the origin, and their distances to it strictly decrease from $\bm{x}_1$ to $\bm{x}_4$, so \ABM\ sets $A = \{\bm{x}_1, \bm{x}_3\}$ and $B = \{\bm{x}_2, \bm{x}_4\}$. Both perfect matchings between $A$ and $B$ have an objective value of $4 + O(\varepsilon)$, whereas the matching $\{\bm{x}_1, \bm{x}_3\}, \{\bm{x}_2, \bm{x}_4\}$ has an objective value of $8 - O(\varepsilon)$. The ratio therefore tends to $1/2$ as $\varepsilon \to 0$.
 
\paragraph{Scope} Lemma~\ref{lem:lower} relies on the decomposition of squared Euclidean distances around a centroid, so the guarantee is stated for squared Euclidean distances only.

\end{document}